\documentclass[11pt,a4paper]{article}
\usepackage[T1]{fontenc}
\usepackage[utf8]{inputenc}
\usepackage[margin=2.3cm]{geometry}
\usepackage{amsmath,amssymb}
\usepackage{newtxtext,newtxmath}
\usepackage{graphicx,bm,booktabs,array,placeins}
\usepackage[numbers,square,sort&compress]{natbib}
\usepackage{authblk}
\usepackage{caption}
\usepackage{url}
\usepackage[hidelinks]{hyperref}
\DeclareRobustCommand{\doi}[1]{doi~\href{https://doi.org/#1}{\nolinkurl{#1}}}
\hypersetup{
 pdftitle={Gravitational-wave responses and noise correlations in geometric time-delay interferometry},
 pdfauthor={Rui Luo, Pan-Pan Wang, Lin-Lin Yang, Xin-Lei Zhao, Cheng-Gang Shao},
 pdfsubject={Gravitational-wave responses and correlated instrumental noise},
 pdfkeywords={gravitational waves, time-delay interferometry, space interferometry}
}
\title{Gravitational-wave responses and noise correlations in geometric time-delay interferometry}
\author[1]{Rui Luo}
\author[2]{Pan-Pan Wang\thanks{Corresponding author, \href{mailto:ppwang@cqu.edu.cn}{ppwang@cqu.edu.cn}}}
\author[1]{Lin-Lin Yang}
\author[1]{Xin-Lei Zhao}
\author[3]{Cheng-Gang Shao}
\affil[1]{National Gravitation Laboratory, MOE Key Laboratory of Fundamental Physical Quantities Measurement, and School of Physics, Huazhong University of Science and Technology, Wuhan 430074, People's Republic of China}
\affil[2]{College of Physics, Chongqing University, Chongqing 401331, China}
\affil[3]{School of Physics and Optoelectronic, Yangtze University, Jingzhou 434023, China}
\date{}
\begin{document}
\maketitle
\begin{abstract}
Space-based gravitational-wave observations will probe the formation and growth of massive black holes and the evolution of compact binaries in the millihertz band. We develop a geometric-link representation of time-delay interferometry using six round-trip and six non-round-trip two-link blocks. Starting from the one-way Doppler response, we derive the elementary block responses and a dictionary that expresses 45 known second-generation combinations in terms of these blocks and residual one-way measurements. Representative Michelson and Relay observables admit compact forms, from which analytical sky-averaged tensor responses are obtained for a fixed, equal-arm constellation. We also derive the noise covariance of the blocks, retaining cross-spectra between blocks that share measurement streams. The resulting noise spectra agree with direct one-way assembly. We apply this formulation to a narrowband compact-binary signal in an equal-arm LISA instrument model. For the Relay example, neglecting the shared-link cross-spectrum changes the low-frequency noise scaling and overestimates the conditional Fisher uncertainties in amplitude or distance by factors of 10.71 at 1~mHz and 3.37 at 3~mHz.
\end{abstract}

\section{Introduction}
Space-based observations of gravitational waves, hereafter GWs, will extend astronomy into the millihertz band. Massive black-hole binaries will probe the assembly and growth of massive black holes \citep{Klein2016MBH}, Galactic compact binaries will constrain the formation and evolution of compact stellar systems \citep{Nelemans2001Binaries,Korol2017Binaries}, and extreme-mass-ratio inspirals will probe the spacetime geometry near massive black holes \citep{Babak2017EMRI}. These observations complement those of stellar-mass mergers by ground-based detectors \citep{LIGO-2016-01,LIGO-2016-02}. The space-based missions LISA \citep{LISA-2024-redbook}, Taiji \citep{taiji-2017-01}, and TianQin \citep{TianQin-2016-01} will measure low-frequency GWs through laser links between freely falling test masses on separate spacecraft. In each link, the received beam is compared with a local laser to measure the GW-induced change in the optical path \citep{tdi-2021-tinto-LRR,unitmodelsim-2023}.

These phase measurements contain both the GW signal and the frequency fluctuations of the transmitting and receiving lasers. The laser fluctuations exceed the expected GW signal by several orders of magnitude and must be suppressed before the astrophysical signal can be recovered \citep{tdi-2021-tinto-LRR}. In an equal-arm interferometer, the same laser fluctuations experience identical propagation delays and cancel when the two beams are compared. The unequal and time-dependent arm lengths of a space-based constellation prevent this direct cancellation. Time-delay interferometry, abbreviated as TDI, restores the cancellation by forming linear combinations of measurements evaluated at appropriately delayed times \citep{tdi-nullchaneel-1999-Armstrong,tdi-2000-Estabrook,tdi-2021-tinto-LRR}. Second-generation combinations account for arm-length variations by retaining the order of the delay operators. Geometric constructions trace the associated optical paths \citep{geometric-tdi-2005-Michele-PRD,geometric-tdi-2022-Wang-PRD}, while algebraic approaches determine combinations that satisfy the laser-noise cancellation conditions \citep{Algebraic-tdi-2002-Dhurandhar,algebraic-TDI-2023-Wu-hust-PRD}.

After laser-noise suppression, the sensitivity of a selected combination is determined by its GW response and the remaining instrumental noise. The response maps a source waveform into the measured data, while the noise spectrum sets the weight of that signal in a signal-to-noise ratio or a parameter uncertainty \citep{Robson2019Sensitivity,LISA-SNR-2021-Babak}. Both follow from the delay coefficients defining the same combination. If measurements are regrouped into optical-path blocks, the signal is assembled coherently and the noise must include cross-spectra between blocks that reuse the same measurements. This relation is particularly relevant to two-link blocks, which need not be independent even when their optical paths are drawn separately.

The one-way Doppler response provides the starting point for calculating the GW signal in such combinations \citep{senstivity-1975-Estabrook-GRG,senstivity-2003-Cornish-PRD,senstivity-2004-Rubbo-PRD}. Numerical and analytical studies have established response and sensitivity functions for different interferometric configurations and polarization states \citep{senstivity-2000-Larson-PRD,senstivity-2019-Liang-PRD,senstivity-2019-Lu-PRD,senstivity-2020-Zhang-PRD}. General analytical formulas for arbitrary combinations were derived in Refs.~\citep{senstivity-2021-Wang-01,senstivity-2021-Wang-02}. A decomposition in terms of lower-generation generators was subsequently used to classify the sensitivities of 45 geometric second-generation combinations \citep{geometric-tdi-2023-Wang-PRD}. More recently, an inverse light-path formulation yielded analytical equal-arm responses, noise spectra, and sensitivity curves for second-generation combinations and their orthogonal channels \citep{Zhang2026SecondGeneration}.

In this work, we use six round-trip and six non-round-trip two-link blocks to display the optical-path structure of the response functions. We derive their relation to the one-way measurement streams and use this dictionary to represent the 45 known second-generation combinations. The mapping preserves the ordered delay operators, with residual one-way terms retained where necessary. For a fixed, equal-arm constellation, the elementary responses give analytical sky-averaged tensor responses for representative Michelson and Relay observables. We further propagate the measurement-noise covariance through the block representation and examine its effect on conditional amplitude or distance uncertainties for a narrowband compact-binary signal. The Relay example shows explicitly how a shared-link cross-spectrum affects the low-frequency noise and the resulting Fisher uncertainties.

The paper is organized as follows. Section~\ref{sec2} introduces the detector configuration and measurement model. Section~\ref{sec3} derives the elementary link responses. Section~\ref{sec4} presents the mapping to second-generation combinations and the analytical equal-arm examples. Section~\ref{sec:binary_application} examines correlated instrumental noise and its effect on compact-binary amplitude measurements. Section~\ref{sec5} summarizes the results. The complete catalogue is given in Appendix~\ref{appendixA}.

\section{Detector configuration and measurement model}\label{sec2}
\subsection{Configuration and notation}

Space-borne GW detectors use laser interferometry to measure changes in the separation of freely falling test masses, denoted by TMs.
Each inter-spacecraft measurement records the heterodyne phase between the received light and a local laser.
GWs perturb the optical paths and hence the measured phase, whose time derivative gives the corresponding frequency shift.

\begin{figure}[tbp]
	\centering
	\includegraphics[width=0.95\textwidth]{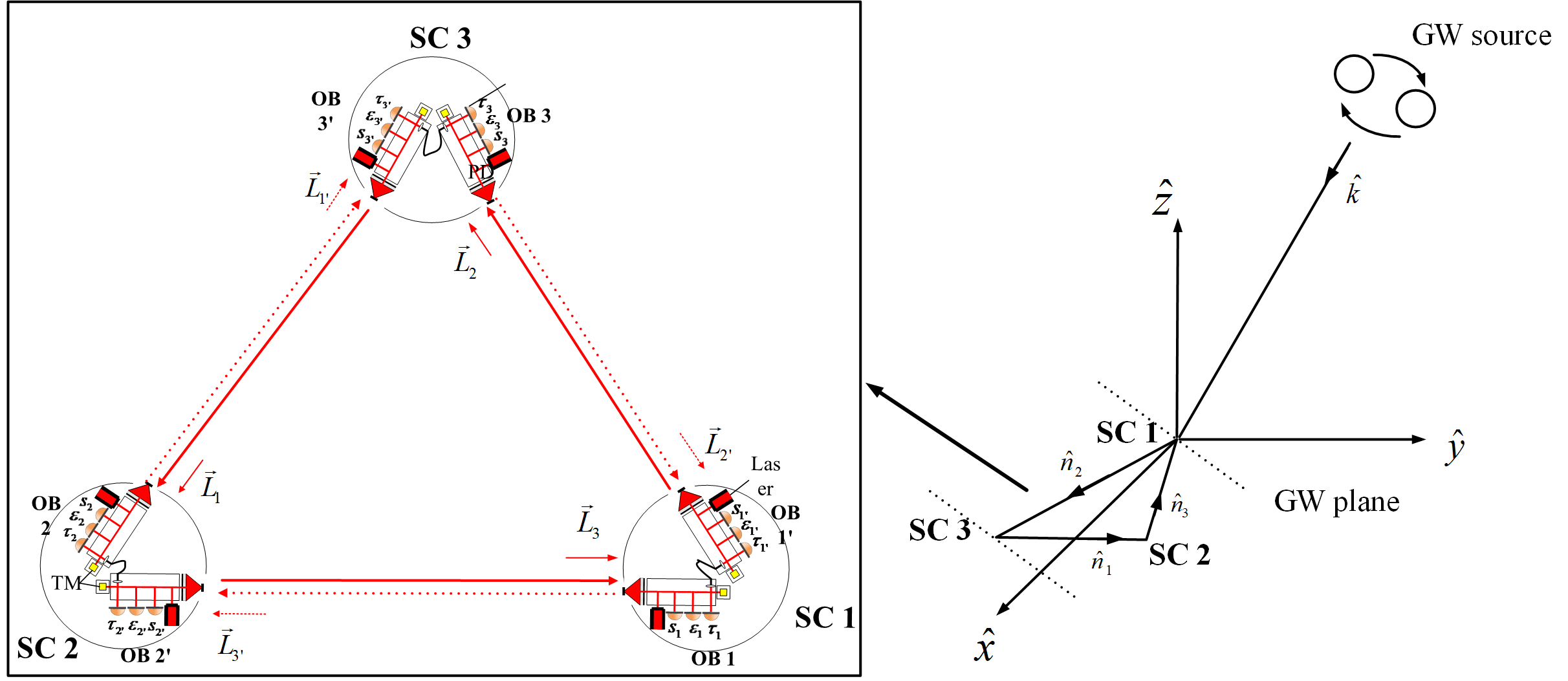}
	\caption{Triangular constellation and directed-link conventions for a space-borne GW detector.}
	\label{fig1}
\end{figure}

The detector consists of three spacecraft, indexed by $i,~i=1,2,3$, that form an approximately equilateral triangle and exchange laser beams, as shown in Figure~\ref{fig1}.
The directed arm lengths $L_{i}$ and $L_{i'}$ describe propagation from spacecraft $i-1$ to spacecraft $i+1$ and in the reverse direction, respectively.
Spacecraft indices follow the cyclic order $1\rightarrow2\rightarrow3\rightarrow1$.

Each spacecraft carries two optical benches, labelled $i$ and $i'$.
The three interferometers on each bench are the inter-satellite interferometer, the test-mass interferometer, and the reference interferometer, abbreviated as ISI, TMI, and RFI, respectively. Their phase measurements are denoted by $s$, $\varepsilon$, and $\tau$.
The measurement model retains the GW signal, laser phase noise, TM motion, and optical metrology system noise, hereafter OMS noise.
Clock noise is assumed to be calibrated, as discussed below.
The phase measurements follow the split-interferometry model \citep{Otto2012Split,unitmodelsim-2023}, specialized to the retained noise terms and the directed-link convention of Figure~\ref{fig1}. On bench $i$ they are
\begin{equation}
	\begin{split}
		&s_i(t) = h_i(t) + D_{i-1} p_{(i+1)'}(t) - p_i(t) + N_i^{\text{OMS}}(t), \\
		&\varepsilon_i(t) = p_{i'}(t) - p_{i}(t) - 2\vec{n}_{(i-1)'} \cdot \vec{\delta}_i(t), \\
		&\tau_i(t) =  p_{i'}(t) - p_i(t),
		\label{data stream in OBi}
	\end{split}
\end{equation}
and
\begin{equation}
	\begin{split}
		&s_{i'}(t) = h_{i'}(t) + D_{(i+1)'} p_{i-1}(t) - p_{i'}(t) + N_{i'}^{\text{OMS}}(t), \\
		&\varepsilon_{i'}(t) = p_i(t) - p_{i'}(t)- 2\vec{n}_{i+1} \cdot \vec{\delta}_{i'}(t), \\
		&\tau_{i'}(t) =  p_i(t) - p_{i'}(t),
		\label{data stream in OBi'}
	\end{split}
\end{equation}
where $h_{i}(t)$ and $h_{i'}(t)$ are the GW contributions, and $p_{i}(t)$ and $p_{i'}(t)$ are the laser phase fluctuations.
The TMI displacement terms include the factor of two from reflection at the test mass. A common optical phase calibration is absorbed into the displacement variables in this phase-level model. The vectors $\vec{\delta}_{i}(t)$ and $\vec{\delta}_{i'}(t)$ describe TM motion driven by acceleration noise; the corresponding unit vectors project this motion onto the measurement axes.
The terms $N_{i}^{\text{OMS}}(t)$ and $N_{i'}^{\text{OMS}}(t)$ denote OMS noise in the ISI measurements.
The nominal arm axes used in the noise projections satisfy $\vec n_{j\prime}=-\vec n_j$. The time-delay operator $D_{i}$ acts on a data stream $f(t)$ as \citep{tdi-2021-tinto-LRR}
\begin{equation}
	\begin{split}
		&D_if(t) = f(t-\frac{L_i(t)}{c}),
		\label{delay operator}
	\end{split}
\end{equation}
where $L_i(t)/c$ is the light travel time and $c$ is the speed of light.
We set $c=1$ unless otherwise stated.
For time-dependent arms, successive delays are evaluated at the corresponding retarded times \citep{tdi-2021-tinto-LRR,geometric-tdi-2005-Michele-PRD}. We abbreviate their ordered products as
\begin{equation}
	\begin{split}
		& D_{ij}f(t) \equiv D_iD_jf(t) = f(t-L_i(t)-L_j(t-L_i(t))).
		\label{compress delay operator}
	\end{split}
\end{equation}
A barred arm index in the catalogue denotes the formal inverse delay operator, $D_{\bar\imath}\equiv D_i^{-1}$. For time-dependent arms, the inverse is defined by composition with $D_i$, and the order of operators is retained in all time-domain products.

Following the standard reduction from bench measurements to one-way TDI inputs \citep{tdi-2021-tinto-LRR,Otto2012Split}, we first combine the ISI, TMI, and RFI data into $\xi_{i}$ and $\xi_{i'}$
\begin{equation}
	\begin{split}
		&\xi_i(t) \equiv s_i(t) - \frac{\varepsilon_i(t) - \tau_i(t)}{2}
          - \frac{D_{i-1}[\varepsilon_{(i+1)'}(t)-\tau_{(i+1)'}(t)]}{2},\\
        &\xi_{i'}(t) \equiv s_{i'}(t) - \frac{\varepsilon_{i'}(t)-\tau_{i'}(t)}{2}
          - \frac{D_{(i+1)'}[\varepsilon_{i-1}(t)-\tau_{i-1}(t)]}{2}.
		\label{xi data}
	\end{split}
\end{equation}
To eliminate the primed laser-noise terms, we introduce
\begin{equation}
	\begin{split}
		z_{i} \equiv \frac{\tau_i - \tau_{i'}}{2}.
		\label{z data}
	\end{split}
\end{equation}
and construct the one-way observables
\begin{equation}
	\begin{split}
		\eta_i(t) &\equiv \xi_i(t) - {D}_{i-1} z_{i+1}, \\
		\eta_{i'}(t) &\equiv \xi_{i'}(t) + z_{i}.
		\label{eta data}
	\end{split}
\end{equation}
The resulting one-way observables have the standard form used in the response and noise analysis of \citet{senstivity-2021-Wang-01}
\begin{equation}
	\begin{split}
		\eta_i(t) &= h_i(t) + {D}_{i-1} p_{i+1}(t) - p_i(t)  \\
		&+  \vec{n}_{i-1} \left[ {D}_{i-1} \vec{\delta}_{(i+1)'}(t) - \vec{\delta}_i(t) \right]  +  N_i^{\text{OMS}}(t),\\
		\eta_{i'}(t) &= h_{i'}(t) + {D}_{(i+1)'} p_{i-1}(t) - p_i(t) \\
		&+  \vec{n}_{i+1} \left[ \vec{\delta}_{i'}(t) -{D}_{(i+1)'} \vec{\delta}_{i-1}(t)  \right]  +  N_{i'}^{\text{OMS}}(t).
		\label{eta data i and i'}
	\end{split}
\end{equation}
The resulting six data streams contain only three independent laser-noise terms, $p_i$, and provide the inputs to the TDI combinations.

\subsection{Instrumental noise model}\label{sec:instrument_noise}
Eq.~\eqref{eta data i and i'} identifies the noise terms that accompany the GW signal in each one-way stream. The remaining laser fluctuations are removed by TDI, whereas TM acceleration noise and OMS noise set the instrumental noise floor. We now specify their spectra to connect the measurement model with the response and noise calculations below.
Laser noise dominates the raw measurements and must be suppressed by TDI.
For the laser-frequency-noise power spectral density, abbreviated as PSD, we retain the white-noise level of the LISA instrument model \citep{unitmodelsim-2023}
\begin{equation}
	\begin{split}
		S_{p}(f) &=\left(30\frac{\mathrm{Hz}}{\sqrt{\mathrm{Hz}}}\right)^2,\\
	\end{split}
	\label{laser noise}
\end{equation}
The low-frequency enhancement of the laser-noise spectrum is not needed here because we assume ideal laser-noise cancellation. We also assume ideal clock-noise calibration using sideband measurements \citep{Otto2012Split}.
After laser and clock noise have been suppressed, TM acceleration noise and OMS noise determine the instrumental sensitivity.
Because of the low received optical power over the inter-spacecraft baselines, we retain OMS noise in the ISI and neglect the corresponding contributions from the local TMI and RFI measurements.
The LISA acceleration-noise and OMS-noise PSDs are adopted from \citet{LISA-SNR-2021-Babak}
\begin{equation}
	\begin{split}
		S_{\text{acc}}(f) &=\left(3~\frac{\mathrm{fm}}{\mathrm{s}^2\sqrt{\mathrm{Hz}}}\right)^2  \times\left[1+(\frac{0.4\mathrm{mHz}}{f})^2\right] \left[1+(\frac{f}{8\mathrm{mHz}})^4\right],\\
	\end{split}
	\label{frequency-dependent acc noise}
\end{equation}
and
\begin{equation}
	\begin{split}
		S_{\text{OMS}}(f)&=\left(15\frac{\mathrm{pm}}{\sqrt{\mathrm{Hz}}}\right)^2\times\left[1+(\frac{2\mathrm{mHz}}{f})^4\right].
	\end{split}
	\label{frequency-dependent oms noise}
\end{equation}
The acceleration amplitude spectral density is converted to an equivalent displacement amplitude spectral density through
\begin{equation}
	\begin{split}
		\frac{\mathrm{m}}{\sqrt{\mathrm{Hz}}}=\frac{1}{\left(2\pi f\right)^{2}}\times \frac{\mathrm{m}}{\mathrm{s}^2\sqrt{\mathrm{Hz}}} .
	\end{split}
	\label{displacement acc noise}
\end{equation}

\subsection{Time-delay interferometry}\label{sec2.3}
The laser-noise term identified above cannot be removed by subtracting simultaneous measurements when the arms are unequal. TDI instead combines measurements with delays matched to the light travel times \citep{tdi-nullchaneel-1999-Armstrong,tdi-2021-tinto-LRR}.
The delayed combinations synthesize optical paths with matching laser-noise contributions, extending the cancellation principle of an equal-arm interferometer to unequal and time-dependent arms.

\begin{figure}[tbp]
	\centering
	\includegraphics[width=0.70\textwidth]{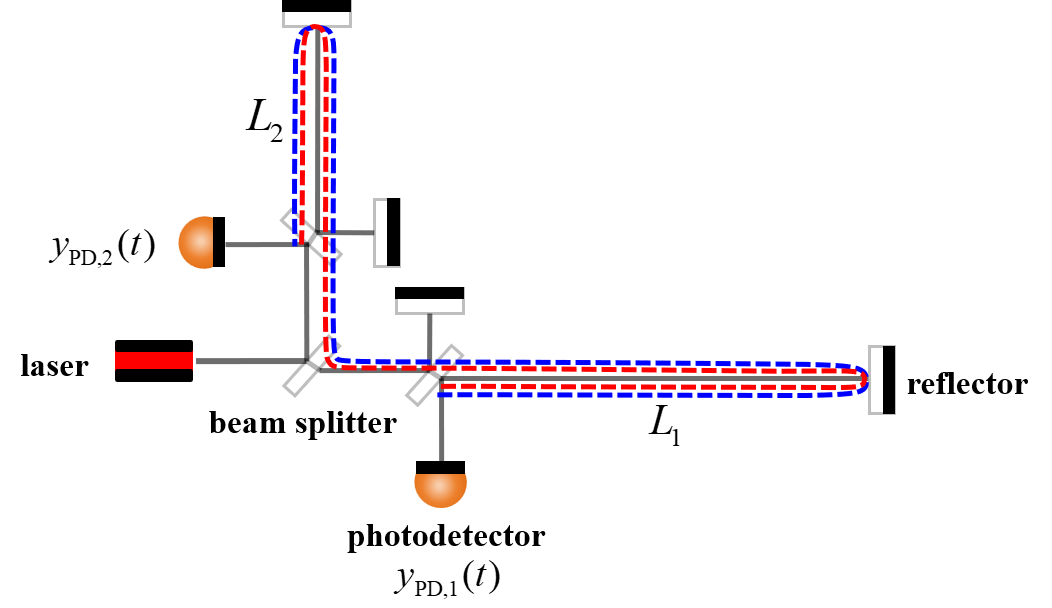}
	\caption{Unequal-arm Michelson interferometer used to illustrate time-delay cancellation. The red and blue paths represent the synthesized paths entering the TDI combination.}
	\label{fig2}
\end{figure}

Consider the unequal-arm Michelson example in Figure~\ref{fig2} with constant arm lengths $L_{1}$ and $L_{2}$.
Following the unequal-arm cancellation construction of \citet{tdi-2021-tinto-LRR}, the two round-trip phase measurements are
\begin{equation}
	\begin{split}
		y_{\text{PD},1}(t) &= h_{1}(t)+p(t-\frac{2L_{1}}{c})-p(t),\\
		y_{\text{PD},2}(t) &= h_{2}(t)+p(t-\frac{2L_{2}}{c})-p(t),
		\label{unequal-arm Michelson}
	\end{split}
\end{equation}
where $h_{i}(t)$ is the GW contribution and $p(t)$ is the common laser phase noise.
Their difference is
\begin{equation}
	\begin{split}
		y_{\text{PD},1}(t)-y_{\text{PD},2}(t) &= h_{1}(t)-h_{2}(t) +p(t-\frac{2L_{1}}{c})-p(t-\frac{2L_{2}}{c}).
		\label{y1-y2}
	\end{split}
\end{equation}
Delaying the first measurement by $2L_{2}/c$ and the second by $2L_{1}/c$ gives a second difference with the same residual laser-noise term,
\begin{equation}
	\begin{split}
		&y_{\text{PD},1}(t-\frac{2L_{2}}{c})-y_{\text{PD},2}(t-\frac{2L_{1}}{c}) \\
		&= h_{1}(t-\frac{2L_{2}}{c})-h_{2}(t-\frac{2L_{1}}{c}) +p(t-\frac{2L_{1}}{c})-p(t-\frac{2L_{2}}{c}).
		\label{tdy1-tdy2}
	\end{split}
\end{equation}
Subtracting Eq.~\eqref{y1-y2} from Eq.~\eqref{tdy1-tdy2} therefore cancels laser noise and gives
\begin{equation}
\begin{split}
 X(t)&=\left[y_{\text{PD},1}(t-\frac{2L_{2}}{c})-y_{\text{PD},2}(t-\frac{2L_{1}}{c})\right]
       -\left[y_{\text{PD},1}(t)-y_{\text{PD},2}(t)\right]\\
     &=\left[y_{\text{PD},2}(t)+y_{\text{PD},1}(t-\frac{2L_{2}}{c})\right]
       -\left[y_{\text{PD},1}(t)+y_{\text{PD},2}(t-\frac{2L_{1}}{c})\right].
\label{X0}
\end{split}
\end{equation}
The sums $y_{\text{PD},2}(t)+y_{\text{PD},1}(t-\frac{2L_{2}}{c})$ and $y_{\text{PD},1}(t)+y_{\text{PD},2}(t-\frac{2L_{1}}{c})$ correspond to the blue and red paths in Figure~\ref{fig2}, respectively.
Their equal total travel times provide a geometric interpretation of the cancellation.

For the triangular constellation, the six one-way observables in Eq.~\eqref{eta data i and i'} are combined as
\begin{equation}
	\begin{split}
		\mathrm{TDI} = \sum_{i=1}^{3} \left( P_i \eta_i +P_{i'} \eta_{i^{\prime}} \right),
	\end{split}
	\label{TDI}
\end{equation}
where $P_{i}$ and $P_{i'}$ are polynomials in the delay operators.
Algebraic and geometric methods determine combinations that cancel laser noise under specified assumptions about the constellation motion.
TDI observables are also distinguished by their number of links and by their signal and noise responses.

In the algebraic approach of \citet{Algebraic-tdi-2002-Dhurandhar}, setting the coefficients of the independent laser-noise terms to zero gives
\begin{equation}
	\begin{split}
		P_1 + P_{1'} - P_2 D_{3'} - P_3 D_2 &= 0, \\
		P_2 + P_{2'} - P_3 D_{1'} - P_1 D_3 &= 0, \\
		P_3 + P_{3'} - P_1 D_{2'} - P_2 D_1 &= 0.
	\end{split}
	\label{TDI equation}
\end{equation}
The search for TDI observables is thus expressed as a system of equations for the delay polynomials.

The geometric approach introduced by \citet{geometric-tdi-2005-Michele-PRD} identifies pairs of synthesized light paths with matching propagation times to the required order.
It provides a direct interpretation of laser-noise cancellation and a systematic way to enumerate combinations.
Subsequent work classified the resulting observables~\citep{geometric-tdi-2020-Muratore-CQG,instrumentnoise-tdi-2022-Olaf}.
Here we consider 45 essential second-generation combinations with up to 16 links, identified using a ternary symbolic search, and adopt the naming convention of \citet{geometric-tdi-2022-Wang-PRD}.

\begin{figure}[tbp]
	\centering
	\includegraphics[width=0.95\textwidth]{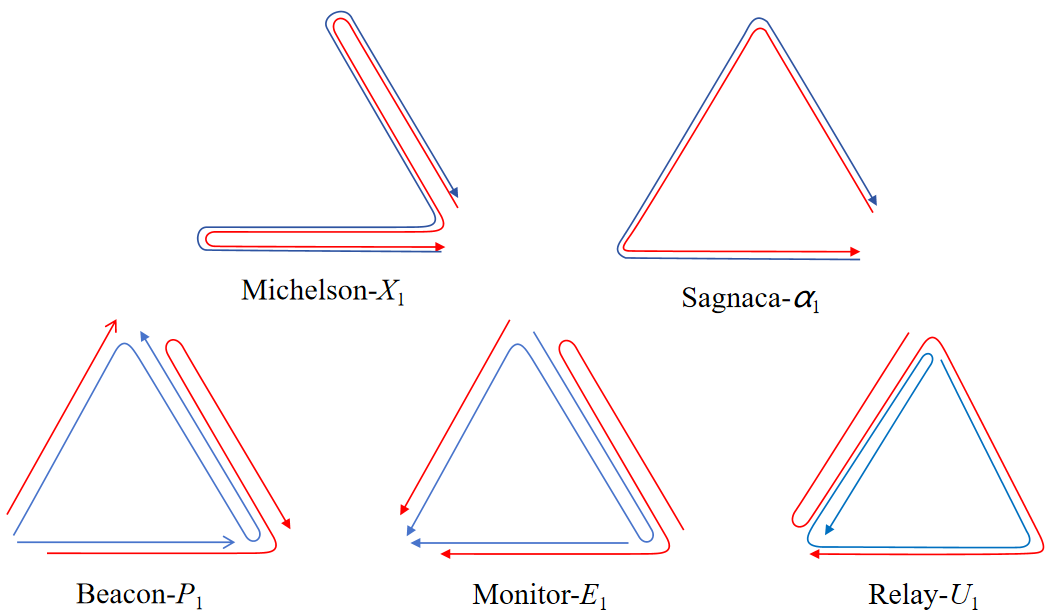}
	\caption{Optical paths for the first-generation Michelson-$X_{1}$, Sagnac-$\alpha_{1}$, Beacon-$P_{1}$, Monitor-$E_{1}$, and Relay-$U_{1}$ TDI combinations. Red and blue indicate the two synthesized paths.}
	\label{fig3}
\end{figure}

The TDI generations correspond to different treatments of constellation motion.
First-generation observables assume constant, reciprocal light travel times; modified first-generation observables allow different travel times in the two propagation directions.
Second-generation constructions account for a rotating and flexing constellation by retaining ordered time delays and matching paths to the relevant order in the arm-length variations.
In the adopted link convention, the standard Michelson-$X_{1}$, Sagnac-$\alpha_{1}$, Beacon-$P_{1}$, Monitor-$E_{1}$, and Relay-$U_{1}$ first-generation combinations are \citep{tdi-2021-tinto-LRR,senstivity-2021-Wang-01}
\begin{equation}
	\begin{split}
		X_1(t) &= (1 - {D}_{33'}) \eta_{1'} + ({D}_{2'} - {D}_{33'2'}) \eta_3 \\
		&+ ({D}_{2'2} - 1) \eta_1 + ({D}_{2'23} - {D}_3) \eta_{2'},\\
		\alpha_1(t) &= \eta_1 - \eta_{1'} + {D}_3 \eta_2 - {D}_{2'1'} \eta_{2'} + {D}_{31} \eta_3 - {D}_{2'} \eta_{3'},\\
		P_1(t) &= {D}_{1} \eta_1 + {D}_{13} \eta_{2'} + {D}_{133'} \eta_{1'} + {D}_{2'} \eta_2 \\
		&- \left( {D}_{1} \eta_{1'} + {D}_{2'} \eta_{2'} + {D}_{2'3} \eta_1 + {D}_{2'33'} \eta_{2} \right),\\
		E_{1}(t) &= \eta_1 + {D}_3 \eta_2 + {D}_{31} \eta_{3'} + {D}_{11'} \eta_{1'} \\
		&- \left( \eta_{1'} + {D}_{2'} \eta_{3'} + {D}_{2'1'} \eta_2 + {D}_{11'} \eta_1 \right),\\
		U_1(t) &= -({D}_3 - {D}_{11'3'}) \eta_{1'} + (1 - {D}_{3'2'1'}) \eta_2 \\
		&- (1 - {D}_{11'}) \eta_{2'} + ({D}_1 - {D}_{3'2'}) \eta_{3'}.
	\end{split}
	\label{first generation TDI}
\end{equation}
Figure~\ref{fig3} shows the corresponding optical paths.
The red and blue paths have matching total light travel times under the first-generation assumptions.

\subsection{Laser locking and frequency planning}\label{sec:locking}
\begin{figure}[tbp]
	\centering
	\includegraphics[width=0.70\textwidth]{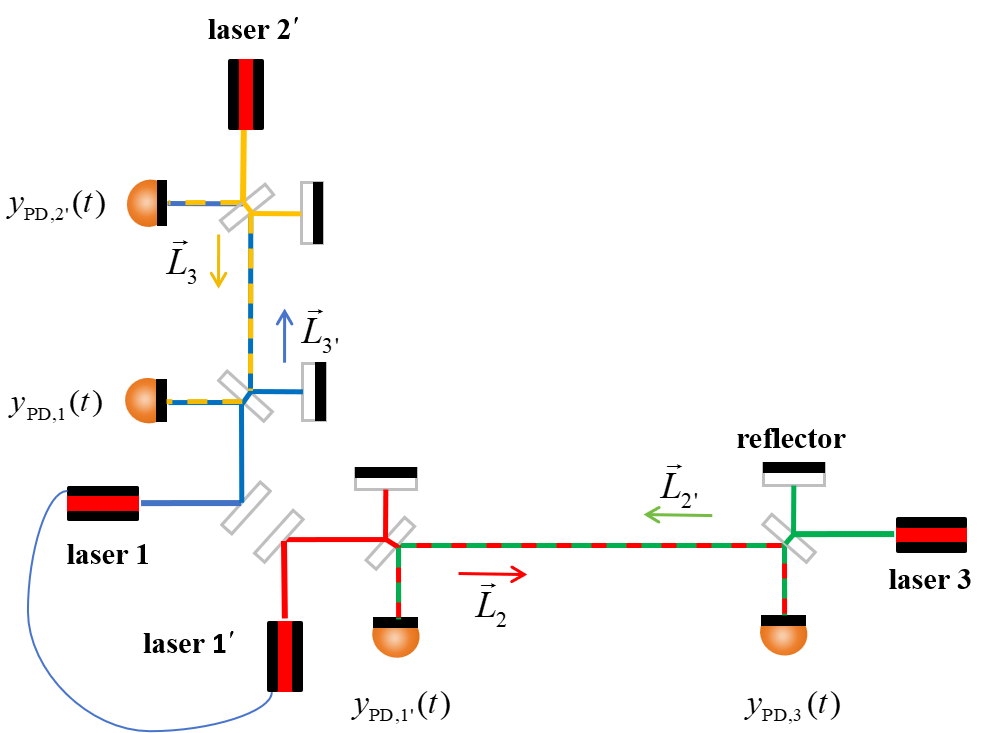}
	\caption{Simplified active laser-link configuration corresponding to an unequal-arm Michelson interferometer. Remote lasers provide the return beams, and phase locking relates their fluctuations to the reference laser.}
	\label{fig4}
\end{figure}

The optical paths in Section~\ref{sec2.3} are synthesized from recorded one-way measurements. Their relation to the physical laser links can be seen by revisiting the Michelson example in Figure~\ref{fig2}. The low received optical power precludes a passive reflected-beam implementation, so the return beam must be generated by a local laser \citep{tdi-2021-tinto-LRR,unitmodelsim-2023}.
The received light can be used as the phase reference for this local laser.
Figure~\ref{fig4} illustrates this arrangement and its relation to the round-trip description used below.
The four phase measurements in this simplified configuration are
\begin{equation}
	\begin{split}
		y_{\text{PD},1}(t) &= h_{1}(t) + p_{2'}(t-\frac{L_{3}}{c}) - p_{1}(t),\\
		y_{\text{PD},1'}(t) &= h_{1'}(t) + p_{3}(t-\frac{L_{2'}}{c}) - p_{1'}(t),\\
		y_{\text{PD},3}(t) &= h_{3}(t) + p_{1'}(t-\frac{L_{2}}{c}) - p_{3}(t),\\
		y_{\text{PD},2'}(t) &= h_{2'}(t) + p_{1}(t-\frac{L_{3'}}{c}) - p_{2'}(t).
		\label{laser lock Michelson}
	\end{split}
\end{equation}
In an idealized locking scheme, laser 1 is the reference.
The measurements at photodetectors $2'$ and 3 are set to zero, and the local reference measurement locks laser $1'$ to laser 1, giving
\begin{equation}
	\begin{split}
		p_{3}(t) &= h_{3}(t) + p_{1'}(t-\frac{L_{2}}{c}),\\
		p_{2'}(t) &= h_{2'}(t) + p_{1}(t-\frac{L_{3'}}{c}),\\
		p_{1}(t)&=p_{1'}(t).
		\label{laser lock}
	\end{split}
\end{equation}
For constant directed arm lengths, we retain the GW and laser-noise contributions and suppress the explicit time arguments. Substituting Eq.~\eqref{laser lock} into Eq.~\eqref{laser lock Michelson} gives
\begin{equation}
\begin{aligned}
 y_{\mathrm{PD},1}&=G_{12}+(A-1)p_1,&
 G_{12}&=h_1+D_3h_{2'},\\
 y_{\mathrm{PD},1'}&=G_{13}+(B-1)p_1,&
 G_{13}&=h_{1'}+D_{2'}h_3,\\
 A&\equiv D_3D_{3'},&B&\equiv D_{2'}D_2.
\end{aligned}
\label{ypd1 and ypd2}
\end{equation}
Applying the opposite round-trip delay to each measurement gives
\begin{equation}
\begin{aligned}
 B y_{\mathrm{PD},1}&=B G_{12}+B(A-1)p_1,\\
 A y_{\mathrm{PD},1'}&=A G_{13}+A(B-1)p_1.
\end{aligned}
\label{delay ypd1 and ypd2}
\end{equation}
For constant arm lengths, $A$ and $B$ commute. The delayed difference that cancels the common laser fluctuation is
\begin{equation}
\begin{aligned}
 X_1^{\mathrm{lock}}
 &=(B-1)y_{\mathrm{PD},1}+(1-A)y_{\mathrm{PD},1'}\\
 &=(B-1)G_{12}+(1-A)G_{13}.
\end{aligned}
\label{X1}
\end{equation}
The laser-noise coefficient is $(B-1)(A-1)+(1-A)(B-1)=0$. Expanding the signal terms recovers the GW contribution to the first-generation Michelson expression in Eq.~\eqref{first generation TDI}. The locking scheme changes how the two round-trip streams are constructed, while their GW contributions remain $G_{12}$ and $G_{13}$.

For the frequency plan considered here, the phase measurement system operates over a heterodyne range of 5--25 MHz.
Relative spacecraft velocities of up to $8\rm{m/s}$ can produce optical Doppler shifts of approximately 10 MHz, so laser-frequency offsets must be selected to keep the beat notes within the readout bandwidth.
The frequency plan uses one reference laser and five phase-locked lasers, with local RFI measurements and remote ISI measurements providing the locking signals.
The six choices of reference laser and six locking schemes give 36 configurations~\citep{unitmodelsim-2023}.
Laser locking controls the beat frequencies and relates the laser fluctuations, while TDI suppresses the residual common laser noise.
We assume ideal frequency control and neglect residual beat-frequency offsets in the response model.

\section{GW responses of geometric links}\label{sec3}
\subsection{Doppler response of a one-way link}

With the one-way measurement streams and their delayed combinations specified, we now evaluate their GW contributions. The response of a TDI observable follows from the delayed sum of the standard one-way Doppler responses \citep{senstivity-1975-Estabrook-GRG,senstivity-2003-Cornish-PRD,senstivity-2004-Rubbo-PRD}. Grouping pairs of measurements into round-trip and non-round-trip geometric links gives an equivalent expression that displays the optical-path structure. We first derive the one-way Doppler response and then obtain the responses of the paired links. The frequency-domain expressions in this section assume fixed spacecraft positions and constant arm lengths.

The measurements in Section~\ref{sec2} are described at the phase level. The GW responses below are expressed as fractional frequency, $y(t)=(2\pi\nu_0)^{-1}\,d\delta\phi(t)/dt$, where $\nu_0$ is the reference optical frequency. For the constant delays used in these response calculations, this conversion commutes with the linear TDI combinations. We retain the symbols $h_i$ for the converted GW contributions when applying the measurement-level mappings to the fractional-frequency responses.

A GW perturbs the light-travel time between spacecraft and hence the frequency measured at the receiving optical bench. We review the derivation to first order in the metric perturbation within the link-response formulation, establishing the notation used for the paired links.

We use the Cartesian coordinate system shown in Figure~\ref{fig1}, with its origin at spacecraft 1. The $x$-axis points from spacecraft 1 towards the midpoint between spacecraft 2 and spacecraft 3, the $y$-axis lies in the detector plane, and the $z$-axis is normal to that plane. The unit vector $\hat{k}$ points from the GW source towards the origin and is written as
\begin{equation}
	\begin{split}
		\hat{k}\equiv - \hat{w} = -\sin\theta \cos\phi \hat{x} - \sin\theta \sin\phi \hat{y} - \cos\theta \hat{z},
		\label{kvector}
	\end{split}
\end{equation}
where $\hat{w}$ points in the opposite direction to $\hat{k}$. The remaining spherical basis vectors are
\begin{equation}
	\begin{split}
		\hat{\theta} &\equiv \partial \hat{w} / \partial \theta = \cos\theta \cos\phi \hat{x} + \cos\theta \sin\phi \hat{y} - \sin\theta \hat{z}, \\
		\hat{\phi}
		&\equiv \partial \hat{w} / (\sin\theta \partial \phi) = -\sin\phi \hat{x} + \cos\phi \hat{y}.
		\label{theta and phi vector}
	\end{split}
\end{equation}

We adopt the Minkowski metric $\eta_{\mu \nu}\equiv \text{diag}(1,-1, -1,-1)$. In the presence of a GW perturbation $h_{\mu\nu}$, the line element becomes
\begin{equation}
	\begin{split}
		ds^{2}= (\eta_{\mu \nu} + h_{\mu \nu})dx^{\mu}dx^{\nu}.
		\label{line element}
	\end{split}
\end{equation}
In general relativity, GWs have two tensor polarization modes, denoted by $h_{+}(t)$ and $h_{\times}(t)$. Using the transverse tensor basis of \citet{senstivity-2004-Rubbo-PRD}, we write the spatial strain as
\begin{equation}
	\begin{split}
		h_{ij}(t) &= h_+(t) \left( \theta_i \theta_j - \phi_i \phi_j \right) + h_\times(t) \left( \theta_i \phi_j + \theta_j \phi_i \right),
		\label{hij}
	\end{split}
\end{equation}
Here $i,j=1,2,3$ are spatial indices. With the adopted metric signature, the spatial strain enters the line element with a minus sign.
The photon trajectory satisfies the null condition
\begin{equation}
	\begin{split}
		0 = c^2 {d}t^2 - {d}x^2 - {d}y^2 - {d}z^2 -h_{ij} dx^i dx^j.
		\label{light ds}
	\end{split}
\end{equation}
For a photon emitted at A and received at B, with spatial coordinates $\vec{x}_{A}$ and $\vec{x}_{B}$, we define
\begin{equation}
	\begin{split}
	L=|\vec{x}_{B}-\vec{x}_{A}|,~	\hat{n}=\frac{\vec{x}_{B}-\vec{x}_{A}}{L}, ~dx^{i}=n^{i}d\lambda,
	\label{n define}
	\end{split}
\end{equation}
where $\lambda$ measures Euclidean distance along the unperturbed path and $\hat{n}$ is its unit direction vector. Eq.~\eqref{light ds} then becomes
\begin{equation}
	\begin{split}
		0 = c^2 \mathrm{d}t^2 - \mathrm{d}\lambda^2 \left( 1 + h_{ij} n^in^j \right).
		\label{ds}
	\end{split}
\end{equation}
The metric perturbation projected along the link is
\begin{equation}
	\begin{split}
	  h(t) &\equiv  h_{ij} n^in^j= h_+(t)\xi_{+} + h_{\times}(t)\xi_{\times},\\
	  \xi_{+}&= (\hat{\theta} \cdot \hat{n})^2 - (\hat{\phi}\cdot\hat{n})^2, \xi_{\times}= 2(\hat{\theta} \cdot \hat{n}) (\hat{\phi}\cdot\hat{n}).
	\label{h+hx}
	\end{split}
\end{equation}

We approximate the GW as a plane wave across the detector, so that the projected perturbation at position $\vec{r}$ is $h ( t - \hat{k} \cdot \vec{r} / c )$. Expanding $1/\sqrt{1+x}\approx 1-\frac{x}{2}$ to first order, Eq.~\eqref{ds} gives
\begin{equation}
	\begin{split}
		d\lambda \approx c \mathrm{d}t \left[ 1 - \frac{1}{2} h \left( t - \hat{k} \cdot \vec{r} / c \right) \right].
		\label{dlambda}
	\end{split}
\end{equation}
Let the emission and reception times be $t_{A}$ and $t_{B}$. Along the unperturbed path, $\vec{r}=\vec{r}_A+c(t-t_A)\hat{n}$, and Eq.~\eqref{dlambda} becomes
\begin{equation}
	\begin{split}
		d\lambda = cdt \left\{ 1 - \frac{1}{2} h \left[ (1 - \hat{k} \cdot \hat{n}) t - \hat{k} \cdot \frac{\vec{r}_A}{c} + t_A \hat{k} \cdot \hat{n} \right] \right\}.
		\label{dlambda2}
	\end{split}
\end{equation}
Integrating along the path yields
\begin{equation}
	\begin{split}
		L = c(t_B - t_A) - \frac{1}{2} \int_{t_A}^{t_B} h \left[ (1 - \hat{k} \cdot \hat{n}) t - \hat{k} \cdot \frac{\vec{r}_A}{c} + t_A \hat{k} \cdot \hat{n} \right] c dt.
		\label{armlength}
	\end{split}
\end{equation}
With the Fourier representation $h(t) = \int d\Omega \tilde{h}(\Omega) e^{-i\Omega t}$, this relation can be written as
\begin{equation}
	\begin{split}
		c(t_B - t_A) &= L + \frac{1}{2} \int_{t_A}^{t_B} \int d\Omega \tilde{h}(\Omega) e^{-i\Omega \left[ (1 - \hat{k} \cdot \hat{n}) t - \hat{k} \cdot \frac{\vec{r}_A}{c} + t_A \hat{k} \cdot \hat{n} \right]} c dt \\
		&\approx L + \frac{1}{2} c \int d\Omega \tilde{h}(\Omega) e^{i\Omega \hat{k} \cdot \frac{\vec{r}_A}{c}} e^{-i\Omega t_B} \frac{e^{\frac{i\Omega L}{c} \hat{k} \cdot \hat{n}} - e^{\frac{i\Omega L}{c}}}{-i\Omega (1 - \hat{k} \cdot \hat{n})},
		\label{t-t0}
	\end{split}
\end{equation}
where the integral is evaluated using the unperturbed light-travel time, $\Omega=2\pi f$, and $f$ is the Fourier frequency. The corresponding beat-note phase is
\begin{equation}
	\begin{split}
		\Delta \phi(t_B) &= -2 \pi \nu_0(t_B-t_A)\\
		&=2 \pi \nu_0 \frac{-L}{c} - \frac{2 \pi \nu_0}{2(1 - \hat{k} \cdot \hat{n})} \times\\
		&\int d\Omega \tilde{h}(\Omega) e^{i\Omega \hat{k} \cdot \frac{\vec{r}_A}{c}} e^{-i\Omega t_B} \frac{e^{\frac{i\Omega L}{c} \hat{k} \cdot \hat{n}} - e^{\frac{i\Omega L}{c}}}{-i\Omega},
		\label{phiAB}
	\end{split}
\end{equation}
where $\nu_0$ is the laser frequency. Differentiating the phase and using $\vec{r}_B=\vec{r}_A+L\hat{n}$ gives the standard one-way fractional Doppler shift \citep{senstivity-1975-Estabrook-GRG,senstivity-2021-Wang-01}
\begin{equation}
	\begin{split}
		h_{B}(t)&=\frac{\delta \nu(t)}{\nu_0} = \frac{1}{2\pi \nu_0} \frac{d\Delta\Phi(t)}{dt}\\
		&= \frac{-1}{2(1 - \hat{k} \cdot \hat{n})} \left[ h\left(t - \hat{k} \cdot \frac{\vec{r}_B}{c}\right) - h\left(t - \hat{k} \cdot \frac{\vec{r}_A}{c} - \frac{L}{c}\right) \right].
		\label{dopplerAB in time}
	\end{split}
\end{equation}
For a time-independent separation $L$, the Fourier transform of Eq.~\eqref{dopplerAB in time} is
\begin{equation}
	\begin{split}
		\tilde{h}_{B}(\Omega)= \frac{\delta \tilde{\nu}(\Omega)}{\nu_0} &= \tilde{h}(\Omega) \frac{ e^{i\Omega \frac{L + \hat{k} \cdot \vec{r}_A}{c}}}{2(1 - \hat{k} \cdot \hat{n})}  \left[ 1 - e^{-i\frac{\Omega L}{c}(1 - \hat{k} \cdot \hat{n})} \right].
		\label{dopplerAB in frequency}
	\end{split}
\end{equation}
Eqs.~\eqref{dopplerAB in time} and \eqref{dopplerAB in frequency} are the time-domain and frequency-domain forms of the same one-way response \citep{senstivity-2021-Wang-01}. Using the notation shown in Figure~\ref{fig1}, light emitted from optical bench $(i+1)'$ and received at optical bench $i$ has propagation distance $L_{i-1}$ and direction $\hat{n}_{i-1}$. Its GW contribution to the measured fractional frequency shift is
\begin{equation}
	\begin{split}
		\tilde{h}_{i}(\Omega) &= \tilde{h}(\Omega)\frac{ e^{i\Omega \frac{L_{i-1} + \hat{k} \cdot \vec{r}_{(i+1)'}}{c}}}{2(1 - \hat{k} \cdot \hat{n}_{i-1})}  \left[ 1 - e^{-i\frac{\Omega L_{i-1}}{c}(1 - \hat{k} \cdot \hat{n}_{i-1})} \right].
		\label{hi}
	\end{split}
\end{equation}
For light emitted from optical bench $i-1$ and received at optical bench $i'$, the distance and direction are $L_{(i+1)'}$ and $\hat{n}_{(i+1)'}=-\hat{n}_{i+1}$. The corresponding GW contribution $h_{i'}$ is
\begin{equation}
	\begin{split}
		\tilde{h}_{i'}(\Omega) &= \tilde{h}(\Omega) \frac{ e^{i\Omega \frac{L_{(i+1)'} + \hat{k} \cdot \vec{r}_{i-1}}{c}}}{2(1 - \hat{k} \cdot \hat{n}_{(i+1)'})}  \left[ 1 - e^{-i\frac{\Omega L_{(i+1)'}}{c}(1 - \hat{k} \cdot \hat{n}_{(i+1)'})} \right].
		\label{hi'}
	\end{split}
\end{equation}
Using the polarization decomposition $\tilde{h}(\Omega)=\tilde{h}_{+}(\Omega)\xi_{+}+\tilde{h}_{\times}(\Omega)\xi_{\times}$ from Eq.~\eqref{h+hx}, Eqs.~\eqref{hi} and \eqref{hi'} take the form
\begin{equation}
	\begin{split}
		\tilde{h}_{i}(\Omega)= F_{h_{i},+}(\Omega)\tilde{h}_{+}(\Omega)+F_{h_{i},\times}(\Omega)\tilde{h}_{\times}(\Omega),
		\label{hi omega}
	\end{split}
\end{equation}
where the polarization response functions for the one-way links are
\begin{equation}
	\begin{split}
		&F_{h_{i},+/\times}(\Omega) \\
		&= \frac{e^{i\Omega (L_{i-1} + \hat{k} \cdot \vec{r}_{(i+1)'})/c}}{2(1 - \hat{k} \cdot \hat{n}_{i-1})} \left[ 1 - e^{-i\Omega L_{i-1} (1 - \hat{k} \cdot \hat{n}_{i-1})/c} \right] \xi_{i-1;+/\times},\\
		&F_{h_{i'},+/\times}(\Omega) \\
		&=\frac{e^{i\Omega (L_{(i+1)'} + \hat{k} \cdot \vec{r}_{i-1})/c}}{2(1 - \hat{k} \cdot \hat{n}_{(i+1)'})} \left[ 1 - e^{-i\Omega L_{(i+1)'} (1 - \hat{k} \cdot \hat{n}_{(i+1)'})/c} \right] \xi_{i+1;+/\times}.
		\label{GW transfer function}
	\end{split}
\end{equation}

The paired-link responses are obtained by delaying and combining these one-way responses.

\subsection{Doppler responses of paired geometric links}
Two adjacent one-way measurements can be grouped into a round-trip link, whose virtual path returns to its starting spacecraft, or a non-round-trip link, whose virtual path ends at the third spacecraft. The six directed one-way links of a triangular constellation define six round-trip and six non-round-trip pairings. These virtual paths are formed by delayed combinations of the measured data streams.

Consider first the virtual round trip that starts at spacecraft $i$, visits spacecraft $i+1$, and returns to spacecraft $i$ at time $t$. Its successive vertices occur at $t-L_{i-1}/c-L_{(i-1)'}/c$, $t-L_{i-1}/c$, and $t$. The earlier one-way response is delayed to the reception time of the second link, giving
\begin{equation}
	\begin{split}
		\tilde{D}_{i-1}h_{(i+1)'} + h_{i}=\sum_{A=+/\times}\left(e^{-i\Omega\frac{L_{i-1}}{c}}F_{h_{(i+1)'},A}\tilde{h}_{A} + F_{h_{i},A}\tilde{h}_{A} \right).
		\label{doppler round-trip 121}
	\end{split}
\end{equation}
For the round trip from spacecraft $i$ through spacecraft $i-1$ and back to spacecraft $i$, the analogous delayed sum is
\begin{equation}
	\begin{split}
		\tilde{D}_{(i+1)'}h_{i-1} + h_{i'}=\sum_{A=+/\times}\left(e^{-i\Omega\frac{L_{(i+1)'}}{c}}F_{h_{i-1},A}\tilde{h}_{A} + F_{h_{(i)'},A}\tilde{h}_{A} \right).
		\label{doppler round-trip 212}
	\end{split}
\end{equation}

\begin{figure}[tbp]
	\centering
	\includegraphics[width=0.72\textwidth]{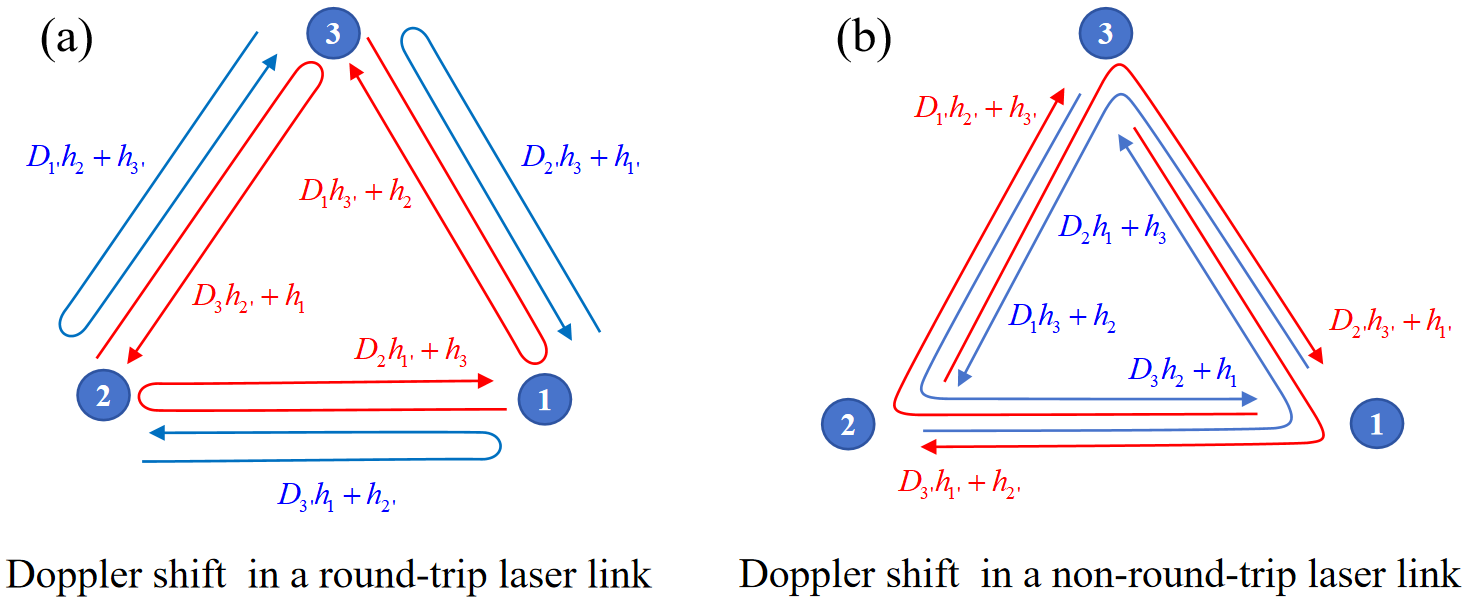}
\caption{Virtual paths of round-trip pairings in (a) and non-round-trip pairings in (b). Each path represents a delayed sum of two one-way measurements.}
	\label{fig5}
\end{figure}

For the non-round-trip path from spacecraft $i$ through spacecraft $i+1$ to spacecraft $i-1$, the paired response is
\begin{equation}
	\begin{split}
		\tilde{D}_{i'}h_{(i+1)'} + h_{(i-1)'}=\sum_{A=+/\times}\left(e^{-i\Omega\frac{L_{i'}}{c}}F_{h_{(i+1)'},A}\tilde{h}_{A} + F_{h_{(i-1)'},A}\tilde{h}_{A} \right).
		\label{doppler non-round-trip 123}
	\end{split}
\end{equation}
For the path from spacecraft $i$ through spacecraft $i-1$ to spacecraft $i+1$, it is
\begin{equation}
	\begin{split}
		\tilde{D}_{i}h_{i-1} + h_{i+1}=\sum_{A=+/\times}\left(e^{-i\Omega\frac{L_{i}}{c}}F_{h_{i-1},A}\tilde{h}_{A} + F_{h_{i+1},A}\tilde{h}_{A} \right).
		\label{doppler non-round-trip 132}
	\end{split}
\end{equation}
Eqs.~\eqref{doppler round-trip 121}--\eqref{doppler non-round-trip 132}, together with cyclic permutations of the spacecraft indices, specify all six round-trip and six non-round-trip pairings and their GW responses. The corresponding virtual paths are shown in Figure~\ref{fig5}(a) and (b), respectively. For time-dependent arm lengths, the pairing is defined by the ordered time-domain delay operators; replacing those operators by frequency-domain phase factors requires the constant-arm approximation used here.

\section{Geometric-link representations of TDI observables}\label{sec4}
\subsection{Mapping rules and representative observables}
The paired responses derived in Section~\ref{sec3} can now be used to assemble complete TDI observables. We apply the same pairings to the one-way data streams, so that each geometric block contains both the GW signal and the corresponding measurement noises. This representation separates the local path geometry from the ordered delay operators that assemble the TDI observable and preserves the signal and noise contributions of the original measurements.

\begin{figure}[tbp]
	\centering
	\includegraphics[width=0.95\textwidth]{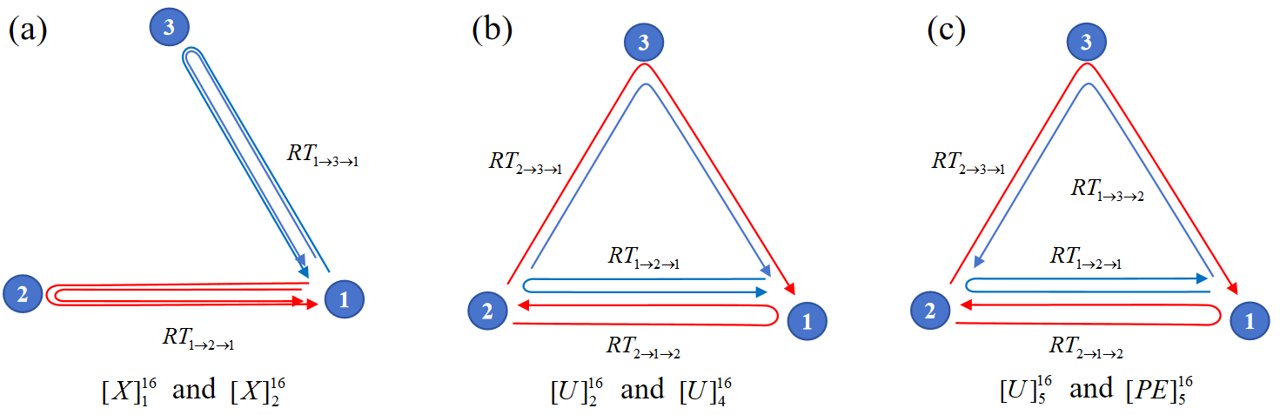}
\caption{Geometric building blocks of $[X]_{1}^{16}$ and $[X]_{2}^{16}$ in (a), $[U]_2^{16}$ and $[U]_4^{16}$ in (b), and $[U]_5^{16}$ and $[PE]_{5}^{16}$ in (c). The paths show the paired-link geometry, and the ordered delay coefficients specify the complete observables.}
	\label{fig6}
\end{figure}

The round-trip mappings are
\begin{equation}
	\begin{split}
		&RT_{3\rightarrow2\rightarrow3}=D_{1'}\eta_{2} + \eta_{3'},RT_{2\rightarrow3\rightarrow2}=D_{1}\eta_{3'} + \eta_{2},\\
		&RT_{1\rightarrow3\rightarrow1}=D_{2'}\eta_{3} + \eta_{1'},RT_{3\rightarrow1\rightarrow3}=D_{2}\eta_{1'} + \eta_{3},\\
		&RT_{2\rightarrow1\rightarrow2}=D_{3'}\eta_{1} + \eta_{2'},RT_{1\rightarrow2\rightarrow1}=D_{3}\eta_{2'} + \eta_{1}.
		\label{rt}
	\end{split}
\end{equation}
The corresponding non-round-trip mappings are
\begin{equation}
	\begin{split}
		&NRT_{3\rightarrow2\rightarrow1}=D_{3}\eta_{2} + \eta_{1},
		NRT_{3\rightarrow1\rightarrow2}=D_{3'}\eta_{1'} + \eta_{2'},\\
		&
		NRT_{2\rightarrow1\rightarrow3}=D_{2}\eta_{1} + \eta_{3},NRT_{2\rightarrow3\rightarrow1}=D_{2'}\eta_{3'} + \eta_{1'},\\
		&NRT_{1\rightarrow3\rightarrow2}=D_{1}\eta_{3} + \eta_{2},
		NRT_{1\rightarrow2\rightarrow3}=D_{1'}\eta_{2'} + \eta_{3'}.
		\label{nrt}
	\end{split}
\end{equation}
We apply Eqs.~\eqref{rt} and \eqref{nrt} to the catalogue of 45 TDI combinations. Some observables can be written entirely in terms of paired links, whereas others retain residual one-way terms, as listed in the appendix. The delay operators remain in their original order, so the regrouping itself does not impose equal or constant arm lengths. For $[X]_{1}^{16}$ and $[X]_{2}^{16}$, this gives
\begin{equation}
	\begin{split}
		[X]_{1}^{16} &=(D_{33'2'22'2}-D_{2'233'})RT_{1\rightarrow2\rightarrow1}+(1-D_{2'2}) RT_{1\rightarrow2\rightarrow1} \\
		&+(D_{33'2'2}-D_{2'233'33'})RT_{1\rightarrow3\rightarrow1}+(D_{33'}-1)RT_{1\rightarrow3\rightarrow1},
	\end{split}
	\label{166.X161}
\end{equation}
and
\begin{equation}
	\begin{split}
		[X]_{2}^{16} &= (1-D_{2'2})RT_{1\rightarrow2\rightarrow1}+(D_{33'2'2}-D_{2'233'\bar{2}\bar{2}'}) RT_{1\rightarrow2\rightarrow1} \\
		&+(D_{33'}-1)RT_{1\rightarrow3\rightarrow1}\\
		&+(D_{2'233'\bar{2}}D_{\bar{2}'}-D_{33'2'233'\bar{2}}D_{\bar{2}'})RT_{1\rightarrow3\rightarrow1}.
	\end{split}
	\label{168.X162}
\end{equation}
Figure~\ref{fig6}(a) shows the two round-trip building blocks shared by $[X]_{1}^{16}$ and $[X]_{2}^{16}$. Both observables have a Michelson-type geometry centred on spacecraft 1, with round trips $RT_{1\rightarrow2\rightarrow1}$ and $RT_{1\rightarrow3\rightarrow1}$. Their different ordered delay coefficients encode how these blocks are combined into the respective second-generation observables.

The analogous representations of $[U]_2^{16}$ and $[U]_4^{16}$ are
\begin{equation}
	\begin{split}
		[U]_2^{16} &=(1-D_{2'1'3'3\bar{1}'\bar{2}'})RT_{1\rightarrow2\rightarrow1} + (D_{33'2'1'}-D_{2'1'})RT_{2\rightarrow1\rightarrow2} \\
		&+ (D_{33'}-1)NRT_{2\rightarrow3\rightarrow1}\\
		&+ (D_{2'1'3'3\bar{1}'}-D_{33'2'1'3'3\bar{1}'})D_{\bar{2}'}NRT_{2\rightarrow3\rightarrow1},
	\end{split}
	\label{161.U162}
\end{equation}
and
\begin{equation}
	\begin{split}
		[U]_4^{16} &=(1+D_{33'2'1'12})RT_{1\rightarrow2\rightarrow1} + (-D_{2'1'}-D_{2'1'3'})RT_{2\rightarrow1\rightarrow2} \\
		&+ (D_{33'}-1)NRT_{2\rightarrow3\rightarrow1} \\
		&+ (D_{33'2'1'}-D_{2'1'3'33'3})NRT_{2\rightarrow3\rightarrow1}.
	\end{split}
	\label{162.U164}
\end{equation}
The building blocks of $[U]_2^{16}$ and $[U]_4^{16}$ are shown in Figure~\ref{fig6}(b). Their local path structure resembles that of the first-generation Relay-$U_{1}$ observable shown in Figure~\ref{fig3}, while their ordered delay coefficients specify the respective second-generation combinations.

For $[U]_5^{16}$ and $[PE]_{5}^{16}$, we obtain
\begin{equation}
	\begin{split}
		[U]_5^{16} &=(1+D_{2'1'3'312\bar{3}'}D_{\bar{3}})RT_{1\rightarrow2\rightarrow1} + (D_{33'}-1)NRT_{2\rightarrow3\rightarrow1} \\
		&- (D_{2'1'}+D_{33'2'1'\bar{3}}D_{\bar{3}'})RT_{2\rightarrow1\rightarrow2}  \\
		& + (D_{33'2'1'\bar{3}\bar{3}'}-D_{2'1'3'3})NRT_{1\rightarrow3\rightarrow2},
	\end{split}
	\label{164.U165}
\end{equation}
and
\begin{equation}
	\begin{split}
		[PE]_{5}^{16} &= (1+D_{2'1'123\bar{3}'}D_{\bar{3}})RT_{1\rightarrow2\rightarrow1}+ (D_{33'}-1)NRT_{2\rightarrow3\rightarrow1} \\
		& - (D_{33'2'1'\bar{3}}+D_{33'2'1'\bar{3}\bar{3}'\bar{3}})D_{\bar{3}'}RT_{2\rightarrow1\rightarrow2} \\
		&+ (D_{2'1'12\bar{3}'\bar{3}\bar{2}}D_{\bar{1}}-D_{2'1'})NRT_{1\rightarrow3\rightarrow2}.
	\end{split}
	\label{1622.PE165}
\end{equation}
Figure~\ref{fig6}(c) shows the building blocks of $[U]_5^{16}$ and $[PE]_{5}^{16}$. Their path geometry can be compared with that of the first-generation Relay-$U_1$, Beacon-$P_{1}$, and Monitor-$E_{1}$ observables. The ordered delay coefficients specify how these paths enter each combination.

These examples identify geometric building blocks shared by first- and second-generation TDI observables. Each regrouped expression retains the laser-noise cancellation of the original ordered-delay combination. Appendix~\ref{appendixA} lists the representations of all 45 combinations, including residual one-way terms where required.

\subsection{Response assembly and shared-link correlations}\label{sec:assembly}
The examples above specify which geometric blocks enter each observable and how they are delayed. To calculate their responses, the complex block amplitudes must be combined before averaging over the sky, as in the standard one-way treatment \citep{senstivity-2021-Wang-01,geometric-tdi-2023-Wang-PRD}. Let $\eta_j$ denote the six one-way streams and let $b_a$ denote a selected round-trip or non-round-trip block. Each mapping in Eqs.~\eqref{rt} and \eqref{nrt} has the form
\begin{equation}
 b_a=\sum_j M_{aj}(D)\eta_j,\qquad
 \mathcal O=\sum_a Q_a(D)b_a+\sum_j r_j(D)\eta_j,
 \label{eq:block_assembly}
\end{equation}
where $M_{aj}$, $Q_a$ and $r_j$ are ordered delay polynomials. The last sum accommodates residual single-link terms. Substitution gives the one-way coefficients $P_j=\sum_a Q_aM_{aj}+r_j$. This identity provides a direct way to compare a grouped expression with its single-link expansion while preserving the order of all delay operators. The blocks form a redundant set of combinations of the six one-way measurements.

For the constant-arm frequency-domain response, collect the selected blocks and any residual one-way streams into a vector $z_\alpha$, with coefficients $w_\alpha(\Omega)$. Linearity gives
\begin{equation}
 F_{\mathcal O,A}(\Omega,\hat k)
 =\sum_\alpha w_\alpha(\Omega)F_{z_\alpha,A}(\Omega,\hat k),
 \qquad A=+,\times.
 \label{eq:response_assembly}
\end{equation}
The sky average must be taken after this coherent sum. With the normalization used below, define
\begin{align}
 \Gamma_{\alpha\beta}(\Omega)
 &=\frac{1}{4\pi}\int d^2\hat k\,
 \sum_{A=+,\times} F_{z_\alpha,A}(\Omega,\hat k)
 F_{z_\beta,A}^{*}(\Omega,\hat k),\label{eq:block_correlation}\\
 \mathcal R_{\mathcal O}(\Omega)
 &=\sum_{\alpha,\beta}w_\alpha(\Omega)w_\beta^{*}(\Omega)
 \Gamma_{\alpha\beta}(\Omega).\label{eq:averaged_assembly}
\end{align}
The off-diagonal terms retain interference between different geometric paths. In particular, the sky-averaged response of a TDI observable cannot in general be obtained by summing the individual block responses. Shared one-way measurements also produce correlated instrumental noise, so their cross-spectra must be included in the noise spectrum of the combined observable.

The paired-link mappings in Eqs.~\eqref{rt} and \eqref{nrt}, together with any residual one-way terms, determine the complex response through Eq.~\eqref{eq:response_assembly}. Eq.~\eqref{eq:averaged_assembly} then gives the sky-averaged response, separating the elementary link responses from the delay coefficients of the selected observable.

\subsection{Analytical response functions under the equal-arm approximation}\label{sec4.2}

We now evaluate the coherent response assembly for a fixed equilateral constellation with equal, constant arm lengths. The one-way response and angular-averaging procedure follow Refs.~\citep{senstivity-2019-Lu-PRD,senstivity-2020-Zhang-PRD,senstivity-2021-Wang-01}; the geometric pairings supply the coefficients for each observable. With the Fourier convention of Eq.~\eqref{dopplerAB in frequency}, the common frequency-domain delay factors are
\begin{equation}
	\begin{split}
		L_i=&L_{i'}=L,\\
		u\equiv& \frac{\Omega L}{c}=\frac{2\pi fL}{c},\\
		\tilde D_i=&\tilde D_{i'}\equiv \tilde D=e^{-iu}.
	\end{split}
	\label{eq:equal_arm_condition_42}
\end{equation}
The GW propagation direction and polarization basis are defined in Eqs.~\eqref{kvector} and \eqref{theta and phi vector}. For a one-way link from spacecraft $a$ to spacecraft $b$, we define
\begin{equation}
	\begin{split}
		\hat n_{ab}=\frac{\vec r_b-\vec r_a}{L},\qquad
		\mu_{ab}=\hat k\cdot \hat n_{ab}.
	\end{split}
	\label{eq:nab_muab}
\end{equation}
The corresponding polarization projection factors follow from Eq.~\eqref{h+hx} as
\begin{equation}
	\begin{split}
		\xi_{ab,+}&=(\hat{\theta}\cdot \hat n_{ab})^2-(\hat{\phi}\cdot \hat n_{ab})^2,\\
		\xi_{ab,\times}&=2(\hat{\theta}\cdot \hat n_{ab})(\hat{\phi}\cdot \hat n_{ab}).
	\end{split}
	\label{eq:xiab_42}
\end{equation}
For an observable $\mathcal O$, we write the GW signal in the frequency domain as
\begin{equation}
	\tilde h_{\mathcal O}(\Omega)=\sum_{A=+,\times}F_{\mathcal O,A}(\Omega)\tilde h_A(\Omega),
	\label{eq:FO_def_42}
\end{equation}
and define its sky-averaged tensor response by
\begin{equation}
	\mathcal R_{\mathcal O}(u)
	=
	\frac{1}{4\pi}\int d^2\hat k
	\left[
	|F_{\mathcal O,+}(u,\hat k)|^2+
	|F_{\mathcal O,\times}(u,\hat k)|^2
	\right].
	\label{eq:sky_average_42}
\end{equation}
The response amplitudes $\sqrt{\mathcal R_{\mathcal O}(u)}$ are obtained by combining the complex link responses before taking the squared modulus and sky average, retaining the interference between different links.

\subsubsection{One-way, round-trip and non-round-trip responses}

Eq.~\eqref{GW transfer function} gives the equal-arm response of the one-way link $a\rightarrow b$ as
\begin{equation}
	\begin{split}
		F_{a\rightarrow b,A}(\Omega)
		=
		\frac{
			e^{i\Omega(L+\hat k\cdot\vec r_a)/c}
		}{
			2(1-\mu_{ab})
		}
		\left[
		1-e^{-iu(1-\mu_{ab})}
		\right]\xi_{ab,A}.
	\end{split}
	\label{eq:single_link_F_42}
\end{equation}
Averaging over propagation directions and summing the two tensor polarizations gives
\begin{equation}
	\mathcal R_{\rm SL}(u)
	=
	\frac{2}{3}
	-\frac{1}{u^2}
	+\frac{\sin 2u}{2u^3}.
	\label{eq:R_single_u_42}
\end{equation}

The round-trip construction in Eqs.~\eqref{doppler round-trip 121} and \eqref{doppler round-trip 212} gives
\begin{equation}
	\begin{split}
		F_{RT_{a\rightarrow b\rightarrow a},A}(\Omega)
		=
		e^{-iu}F_{a\rightarrow b,A}(\Omega)
		+
		F_{b\rightarrow a,A}(\Omega).
	\end{split}
	\label{eq:F_RT_42}
\end{equation}
Using $\hat n_{ba}=-\hat n_{ab}$ and $\xi_{ba,A}=\xi_{ab,A}$, Eq.~\eqref{eq:F_RT_42} can be rewritten as
\begin{equation}
	\begin{split}
		F_{RT_{a\rightarrow b\rightarrow a},A}(\Omega)
		=
		e^{i\Omega\hat k\cdot\vec r_a/c}
		\xi_{ab,A}\,
		\mathcal T_{\rm RT}(u,\mu_{ab}),
	\end{split}
	\label{eq:F_RT_TRT_42}
\end{equation}
where
\begin{equation}
	\mathcal T_{\rm RT}(u,\mu)
	=
	\frac{1-e^{-iu(1-\mu)}}{2(1-\mu)}
	+
	\frac{e^{iu(1+\mu)}-1}{2(1+\mu)}.
	\label{eq:T_RT_42}
\end{equation}
The corresponding sky-averaged response is
\begin{equation}
	\mathcal R_{\rm RT}(u)
	=
	1-\frac{\cos 2u}{3}
	-\frac{3+\cos 2u}{u^2}
	+\frac{2\sin 2u}{u^3}.
	\label{eq:R_RT_u_42}
\end{equation}

Similarly, Eqs.~\eqref{doppler non-round-trip 123} and \eqref{doppler non-round-trip 132} give the response of $NRT_{a\rightarrow b\rightarrow c}$ as
\begin{equation}
	\begin{split}
		F_{NRT_{a\rightarrow b\rightarrow c},A}(\Omega)
		=
		e^{-iu}F_{a\rightarrow b,A}(\Omega)
		+
		F_{b\rightarrow c,A}(\Omega).
	\end{split}
	\label{eq:F_NRT_42}
\end{equation}
For an equilateral triangular constellation, the sky-averaged non-round-trip response can be expressed as
\begin{equation}
	\mathcal R_{\rm NRT}(u)
	=
	f_1(u)+\frac{3}{2}f_3(u),
	\label{eq:R_NRT_f_42}
\end{equation}
where
\begin{equation}
	f_1(u)
	=
	\frac{4}{3}
	-\frac{2}{u^2}
	+\frac{\sin 2u}{u^3},
	\label{eq:f1_42}
\end{equation}
and
\begin{equation}
	\begin{split}
		f_3(u)
		&=
		\ln\frac{4}{3}
		-\frac{5}{18}
		+\frac{-5\sin u+8\sin 2u-3\sin 3u}{8u}  \\
		&\quad
		-\frac{4+9\cos u+12\cos 2u+\cos 3u}{24u^2}  \\
		&\quad
		+\frac{-5\sin u+8\sin 2u+5\sin 3u}{24u^3}  \\
		&\quad
		+\operatorname{Ci}(3u)-2\operatorname{Ci}(2u)+\operatorname{Ci}(u).
	\end{split}
	\label{eq:f3_42}
\end{equation}
Here $\operatorname{Ci}(u)$ is the cosine-integral function.

\begin{figure}[tbp]
	\centering
	\includegraphics[width=0.72\textwidth]{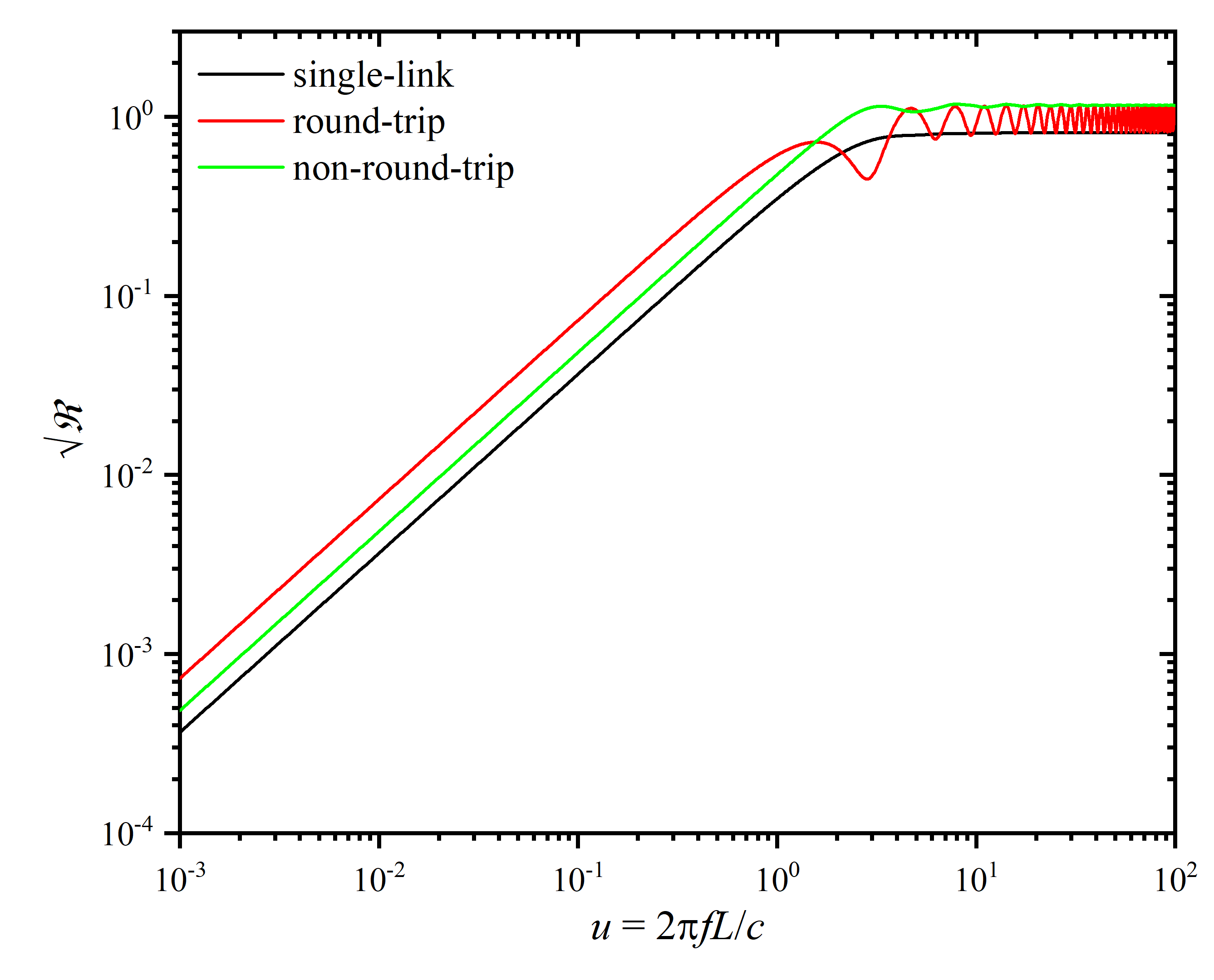}
\caption{Sky-averaged response amplitudes $\sqrt{\mathcal R_{\mathcal O}(u)}$ of the one-way, round-trip, and non-round-trip building blocks for a fixed equal-arm constellation. The analytical responses are given in Eqs.~\eqref{eq:R_single_u_42}, \eqref{eq:R_RT_u_42}, and \eqref{eq:R_NRT_u_42}.}
	\label{fig7}
\end{figure}

Equivalently, Eq.~\eqref{eq:R_NRT_f_42} can be expanded as
\begin{equation}
	\begin{split}
		\mathcal R_{\rm NRT}(u)
		&=
		\frac{11}{12}
		+\frac{3}{2}\ln\frac{4}{3}
		+\frac{-15\sin u+24\sin 2u-9\sin 3u}{16u}  \\
		&\quad
		-\frac{36+9\cos u+12\cos 2u+\cos 3u}{16u^2} \\
		&\quad
		+\frac{-5\sin u+24\sin 2u+5\sin 3u}{16u^3} \\
		&\quad
		+\frac{3}{2}
		\left[
		\operatorname{Ci}(3u)-2\operatorname{Ci}(2u)+\operatorname{Ci}(u)
		\right].
	\end{split}
	\label{eq:R_NRT_u_42}
\end{equation}
In the long-wavelength limit, the leading-order terms are
\begin{equation}
	\begin{split}
		\mathcal R_{\rm SL}(u)&=\frac{2}{15}u^2+\mathcal O(u^4),\\
		\mathcal R_{\rm RT}(u)&=\frac{8}{15}u^2+\mathcal O(u^4),\\
		\mathcal R_{\rm NRT}(u)&=\frac{7}{30}u^2+\mathcal O(u^4).
	\end{split}
	\label{eq:low_freq_geom_42}
\end{equation}
Figure~\ref{fig7} compares the sky-averaged response amplitudes of the one-way, round-trip, and non-round-trip links.

\subsubsection{Response of \texorpdfstring{$[X]_{1}^{16}$}{X16}}

Eq.~\eqref{rt} defines the two round-trip blocks centred on spacecraft 1 as
\begin{equation}
	\begin{split}
		RT_{1\rightarrow2\rightarrow1}=D_3\eta_{2'}+\eta_1,\qquad
		RT_{1\rightarrow3\rightarrow1}=D_{2'}\eta_3+\eta_{1'}.
	\end{split}
	\label{eq:RT121_RT131_42}
\end{equation}
Under the equal-arm condition, Eq.~\eqref{166.X161} reduces to
\begin{equation}
	\begin{split}
		[X]_{1,\rm eq}^{16}
		=
		(1-\tilde D^2)(1-\tilde D^4)
		\left(
		RT_{1\rightarrow2\rightarrow1}
		-
		RT_{1\rightarrow3\rightarrow1}
		\right).
	\end{split}
	\label{eq:X16_equal_form_42}
\end{equation}
Therefore, the response function is
\begin{equation}
	\begin{split}
		F_{[X]_{1}^{16},A}(\Omega)
		=
		(1-e^{-2iu})(1-e^{-4iu})\times \\
		\left[
		F_{RT_{1\rightarrow2\rightarrow1},A}(\Omega)
		-
		F_{RT_{1\rightarrow3\rightarrow1},A}(\Omega)
		\right].
	\end{split}
	\label{eq:F_X16_42}
\end{equation}
Eq.~\eqref{eq:F_X16_42} expresses the second-generation Michelson-type response in terms of two reusable round-trip amplitudes. Their interference is retained when the complete response is squared and averaged.

For comparison, the known analytical response of the first-generation Michelson observable, expressed with the tensor-sum normalization of Eq.~\eqref{eq:sky_average_42}, is \citep{senstivity-2020-Zhang-PRD,senstivity-2021-Wang-01}
\begin{equation}
	\begin{split}
	\mathcal R_{X_1}(u)
	&=
	2\sin^2 u
	\Bigg[
	3+4\ln2
	-6\cos 2u
	\left(
	\ln\frac{4}{3}-\frac{5}{18}
	\right)
	-\frac{8\cos^2u}{3} \\
	&\quad
	-\frac{7\sin u-2\sin2u}{u}
	+\frac{5\cos u-8\cos^2u}{u^2}
	-\frac{5\sin u-4\sin2u}{u^3} \\
	&\quad
	-4\left[\operatorname{Ci}(2u)-\operatorname{Ci}(u)\right] \\
	&\quad
	-6\cos2u
	\left[
	\operatorname{Ci}(3u)-2\operatorname{Ci}(2u)+\operatorname{Ci}(u)
	\right] \\
	&\quad
	-6\sin2u
	\left[
	\operatorname{Si}(3u)-2\operatorname{Si}(2u)+\operatorname{Si}(u)
	\right]
	\Bigg],
	\end{split}
	\label{eq:R_X1_analytic_42}
\end{equation}
where $\operatorname{Si}(u)$ is the sine-integral function. Since Eq.~\eqref{eq:X16_equal_form_42} is equivalent to
\begin{equation}
	[X]_{1,\rm eq}^{16}=-(1-\tilde D^4)X_1,
	\label{eq:X16_X1_relation_42}
\end{equation}
with an overall sign that does not affect the response power, the analytical response of $[X]_{1}^{16}$ is
\begin{equation}
	\mathcal R_{[X]_{1}^{16}}(u)
	=
	|1-e^{-4iu}|^2\mathcal R_{X_1}(u)
	=
	4\sin^2(2u)\mathcal R_{X_1}(u).
	\label{eq:R_X16_analytic_42}
\end{equation}

\subsubsection{Response of \texorpdfstring{$[U]_{2}^{16}$}{U16}}

For the Relay-type second-generation observable $[U]_{2}^{16}$, the equal-arm reduction of Eq.~\eqref{161.U162} is
\begin{equation}
	\begin{split}
		[U]_{2,\rm eq}^{16}
		&=(1-\tilde D^2)RT_{1\rightarrow2\rightarrow1}
		+(\tilde D^4-\tilde D^2)RT_{2\rightarrow1\rightarrow2} \\
		&\quad
		+(\tilde D^2-1)NRT_{2\rightarrow3\rightarrow1}
		+(\tilde D^2-\tilde D^4)NRT_{2\rightarrow3\rightarrow1}.
	\end{split}
	\label{eq:U16_equal_form_42}
\end{equation}
Equivalently, the two non-round-trip terms can be combined as
\begin{equation}
	\begin{split}
		[U]_{2,\rm eq}^{16}
		&=(1-\tilde D^2)RT_{1\rightarrow2\rightarrow1}
		+(\tilde D^4-\tilde D^2)RT_{2\rightarrow1\rightarrow2} \\
		&\quad
		-(1-\tilde D^2)^2NRT_{2\rightarrow3\rightarrow1}.
	\end{split}
	\label{eq:U16_equal_form_compact_42}
\end{equation}
Thus the GW response function is
\begin{equation}
	\begin{split}
		F_{[U]_{2}^{16},A}(\Omega)
		&=(1-e^{-2iu})F_{RT_{1\rightarrow2\rightarrow1},A}(\Omega) \\
		&\quad +(e^{-4iu}-e^{-2iu})F_{RT_{2\rightarrow1\rightarrow2},A}(\Omega) \\
		&\quad +(2e^{-2iu}-1-e^{-4iu})F_{NRT_{2\rightarrow3\rightarrow1},A}(\Omega),
	\end{split}
	\label{eq:F_U16_42}
\end{equation}
where $A=+,\times$. Thus $[U]_{2}^{16}$ is assembled from two round-trip blocks and one non-round-trip block under this approximation. The two non-round-trip terms in Eq.~\eqref{eq:U16_equal_form_42} share the combined coefficient $2e^{-2iu}-1-e^{-4iu}$.

\begin{figure}[tbp]
	\centering
	\includegraphics[width=0.72\textwidth]{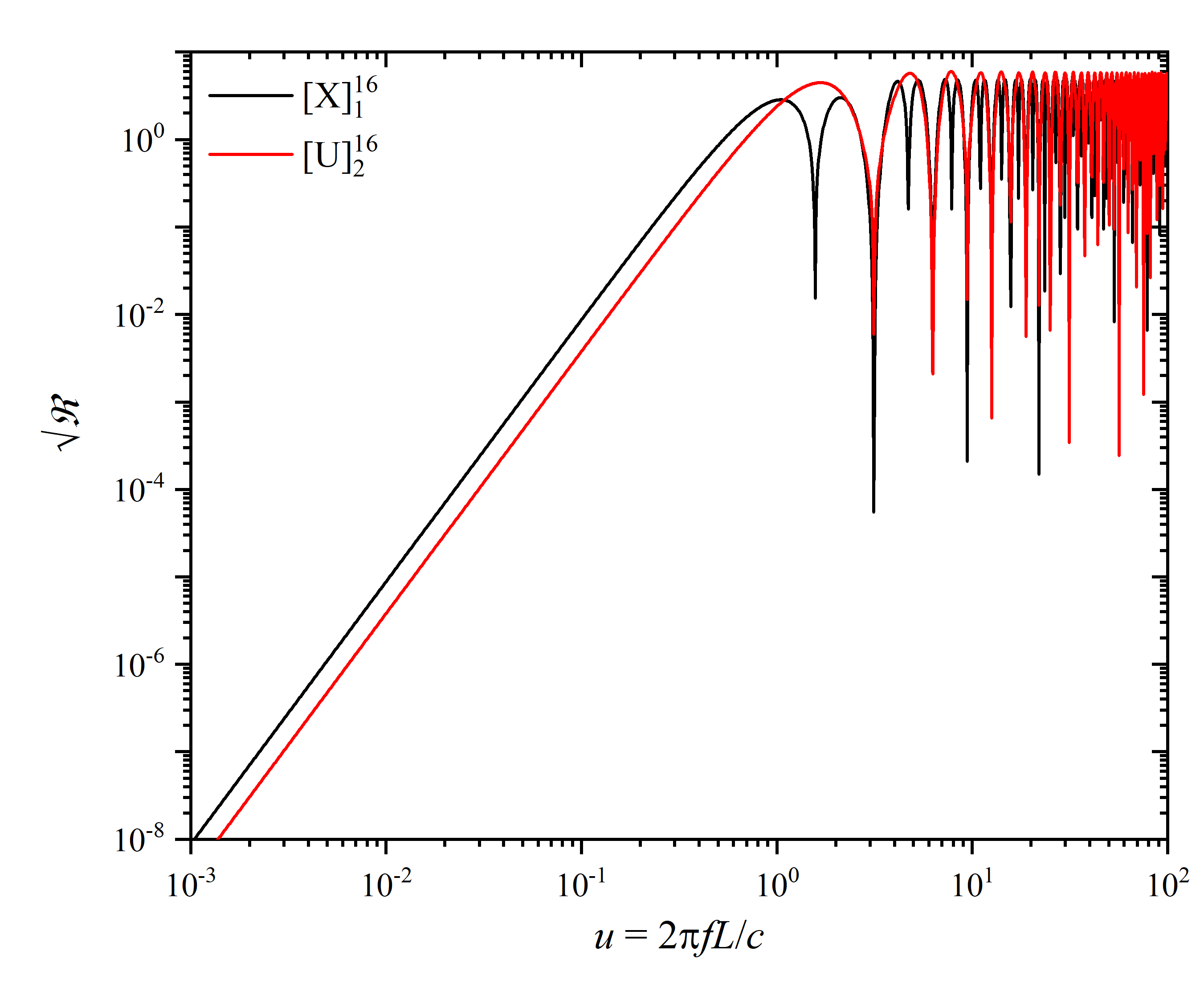}
\caption{Sky-averaged response amplitudes of the second-generation TDI observables $[X]_{1}^{16}$ and $[U]_{2}^{16}$ for a fixed equal-arm constellation. The analytical responses are given in Eqs.~\eqref{eq:R_X16_analytic_42} and \eqref{eq:R_U16_factorized_42}.}
	\label{fig8}
\end{figure}

The analytical sky-averaged response follows from the factorized form for arbitrary TDI combinations~\citep{senstivity-2021-Wang-01},
\begin{equation}
	\begin{split}
	\mathcal R_{[U]_{2}^{16}}(u)
	&=\frac{1}{2}C^{U}_1(u)f_1(u)+C^{U}_2(u)f_2(u) \\
	&\quad +\frac{3}{4}C^{U}_3(u)f_3(u)
	-\frac{3}{4}C^{U}_4(u)f_4(u)
	+\frac{1}{4}C^{U}_5(u)f_5(u).
	\end{split}
	\label{eq:R_U16_factorized_42}
\end{equation}
The functions $f_i(u)$ are analytical response kernels. Alongside $f_1(u)$ and $f_3(u)$ in Eqs.~\eqref{eq:f1_42} and \eqref{eq:f3_42}, the required kernels are
\begin{equation}
	f_2(u)=\frac{-u\cos u+\sin u}{u^3}-\frac{\cos u}{3},
	\label{eq:f2_42}
\end{equation}
\begin{equation}
	\begin{split}
	f_4(u)
	&=\frac{-5\cos u+8\cos2u}{8u}-\frac{3\cos3u}{8u} \\
	&\quad +\frac{9\sin u+12\sin2u}{24u^2}+\frac{\sin3u}{24u^2} \\
	&\quad -\frac{8+5\cos u-8\cos2u}{24u^3}+\frac{5\cos3u}{24u^3} \\
	&\quad +2\operatorname{Si}(2u)-\operatorname{Si}(3u)-\operatorname{Si}(u),
	\end{split}
	\label{eq:f4_42}
\end{equation}
\begin{equation}
	\begin{split}
	f_5(u)
	&=-\ln4+\frac{7}{6}+\frac{11\sin u-4\sin2u}{4u} \\
	&\quad -\frac{10+5\cos u}{4u^2}+\frac{\cos2u}{2u^2} \\
	&\quad +\frac{5\sin u+4\sin2u}{4u^3}
	+2\left[\operatorname{Ci}(2u)-\operatorname{Ci}(u)\right].
	\end{split}
	\label{eq:f5_42}
\end{equation}
Expanding Eq.~\eqref{eq:U16_equal_form_42} in the six one-way data streams gives the coefficients
\begin{equation}
	\begin{split}
	P_1&=1-\tilde D^2-\tilde D^3+\tilde D^5,
	\qquad P_2=0,
	\qquad P_3=0,\\
	P_{1'}&=-1+2\tilde D^2-\tilde D^4,\\
	P_{2'}&=\tilde D-\tilde D^2-\tilde D^3+\tilde D^4,\\
	P_{3'}&=-\tilde D+2\tilde D^3-\tilde D^5.
	\end{split}
	\label{eq:U16_single_link_polynomial_42}
\end{equation}
Substituting Eq.~\eqref{eq:U16_single_link_polynomial_42} into the coefficient definitions gives
\begin{equation}
	\begin{split}
	C^{U}_1(u)&=20-24\cos2u-2\cos3u+4\cos4u+2\cos5u,\\
	C^{U}_2(u)&=4+4\cos u-8\cos2u-4\cos3u+4\cos4u,\\
	C^{U}_3(u)&=24-8\cos u-32\cos2u+12\cos3u \\
	&\quad +8\cos4u-4\cos5u,\\
	C^{U}_4(u)&=-16\sin u-8\sin2u+12\sin3u \\
	&\quad +4\sin4u-4\sin5u,\\
	C^{U}_5(u)&=-12-8\cos u+16\cos2u+12\cos3u \\
	&\quad -4\cos4u-4\cos5u.
	\end{split}
	\label{eq:CU_42}
\end{equation}
Figure~\ref{fig8} compares the sky-averaged response amplitudes of $[X]_1^{16}$ and $[U]_2^{16}$. Their amplitudes and frequency dependence reflect the different delay coefficients and interference between the link responses.

\subsubsection{Response under ideal phase locking}

\begin{figure}[tbp]
	\centering
	\includegraphics[width=0.72\textwidth]{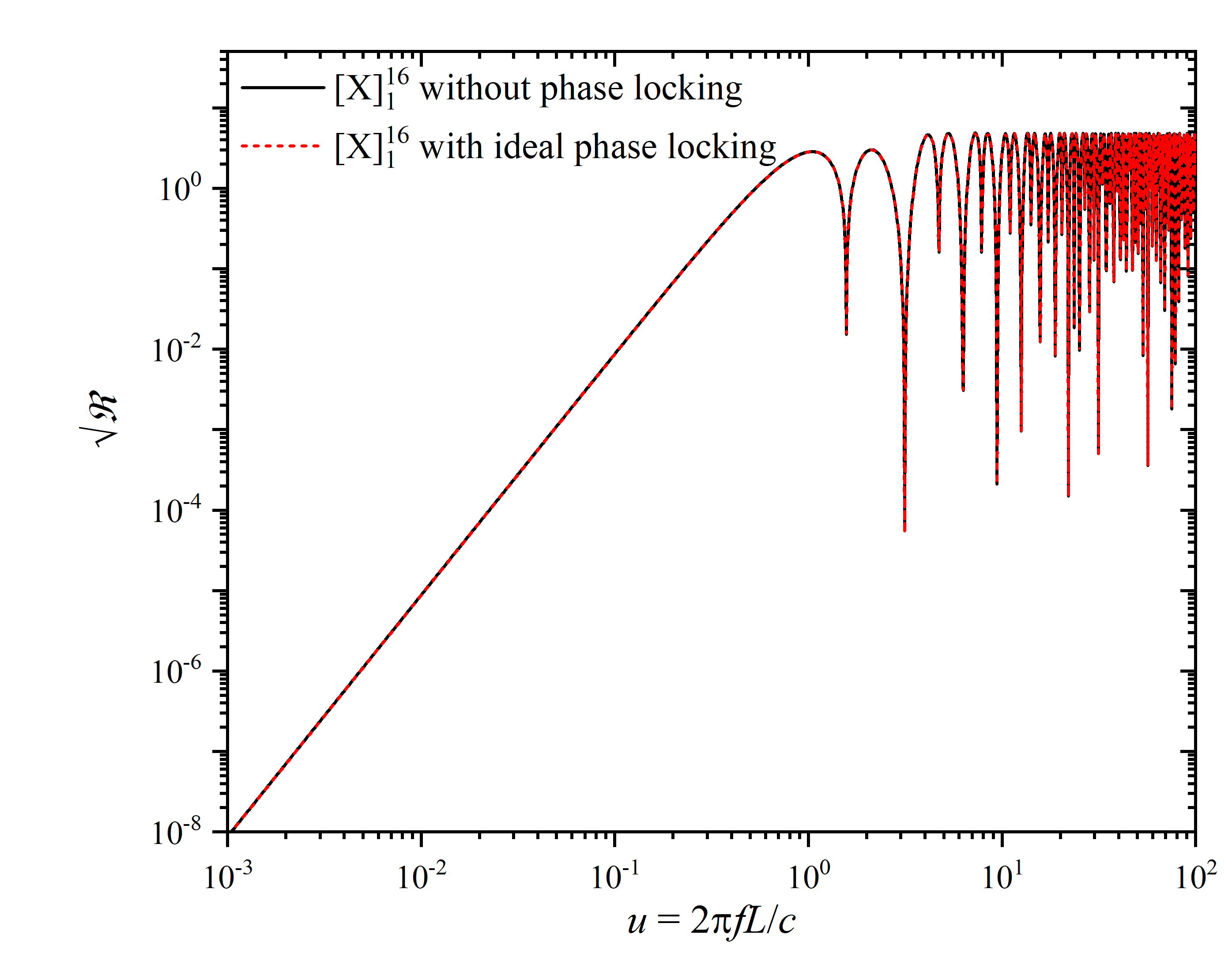}
\caption{Sky-averaged GW response amplitudes of $[X]_{1}^{16}$ with and without ideal phase locking in the equal-arm model. The curves coincide according to Eq.~\eqref{eq:X16_lock_equal_R_42}.}
	\label{fig9}
\end{figure}

The phase-locking scheme in Section~\ref{sec:locking} changes how the phase measurements are acquired and combined. For the simplified Michelson configuration in Eqs.~\eqref{laser lock}--\eqref{X1}, ideal locking preserves the GW contribution to the corresponding TDI observable. Under the same equal-arm assumptions as above, its response is therefore
\begin{equation}
	F^{\rm lock}_{[X]_{1}^{16},A}(\Omega)
	=
	F_{[X]_{1}^{16},A}(\Omega),
	\label{eq:X16_lock_equal_F_42}
\end{equation}
and hence
\begin{equation}
	\mathcal R^{\rm lock}_{[X]_{1}^{16}}(u)
	=
	\mathcal R_{[X]_{1}^{16}}(u).
	\label{eq:X16_lock_equal_R_42}
\end{equation}
The coincident curves shown in Figure~\ref{fig9} illustrate the equality of the GW responses under ideal phase locking, with residual locking noise and clock noise neglected.

\section{Application to conditional compact-binary measurements}
\label{sec:binary_application}
\subsection{Noise covariance of the geometric blocks}
\label{sec:binary_covariance}

We use the fixed, equal-arm configuration of Section~\ref{sec4.2} and the instrumental-noise model of Section~\ref{sec:instrument_noise}. The covariance below includes only the OMS and TM secondary-noise contributions to the fractional-frequency one-way measurements. Laser noise cancels in the constant-arm TDI combination considered here, and clock noise is assumed to be calibrated.

We express the optical metrology and acceleration noise of each test mass as
one-sided fractional-frequency PSDs \citep{LISA-SNR-2021-Babak},
\begin{equation}
 S_{\rm op}(f)=\left(\frac{2\pi f}{c}\right)^2S_{\rm OMS}(f),\qquad
 S_{\rm pm}(f)=\frac{S_{\rm acc}(f)}{(2\pi f c)^2}.
 \label{eq:application_noise_conversion}
\end{equation}
The optical metrology noises in the six directed links are assumed to be independent, as are the acceleration noises of the six test masses. Oppositely directed measurements along the same arm are nevertheless correlated because they contain the same two test-mass noises
\citep{LISA-SNR-2021-Babak}. With the measurement order
$\boldsymbol\eta=(\eta_1,\eta_{1'},\eta_2,\eta_{2'},\eta_3,\eta_{3'})^T$,
their secondary-noise covariance has diagonal elements
$v=S_{\rm op}+2S_{\rm pm}$ and nonzero off-diagonal elements
\begin{equation}
 (C_\eta)_{1,2'}=(C_\eta)_{1',3}=(C_\eta)_{2,3'}
 =2\cos u\,S_{\rm pm},
 \label{eq:application_oneway_covariance}
\end{equation}
together with their Hermitian counterparts. All other off-diagonal elements
vanish under the stated noise assumptions. These real cross spectra follow
from the sum of the delayed and undelayed contributions of the two shared
test masses.

For the Relay observable $[U]_2^{16}$, write $D=e^{-iu}$ and $a=1-D^2$,
and collect its blocks as
\begin{equation}
 \boldsymbol z=
 \left(RT_{1\rightarrow2\rightarrow1},
 RT_{2\rightarrow1\rightarrow2},
 NRT_{2\rightarrow3\rightarrow1}\right)^T.
 \label{eq:application_blocks}
\end{equation}
For this ordering, the block map is $\boldsymbol z=M\boldsymbol\eta$, with
\begin{equation}
 M=\begin{pmatrix}
 1&0&0&D&0&0\\
 D&0&0&1&0&0\\
 0&1&0&0&0&D
 \end{pmatrix}.
 \label{eq:application_block_map}
\end{equation}
The equal-arm expression is $U=\boldsymbol w\boldsymbol z$, with
$\boldsymbol w=(a,-D^2a,-a^2)$. The covariance $C_z=MC_\eta M^\dagger$ is
\begin{equation}
 C_z=\begin{pmatrix}
 S_{\rm RT}&C_{\rm RT}&0\\
 C_{\rm RT}&S_{\rm RT}&0\\
 0&0&S_{\rm NRT}
 \end{pmatrix},
 \label{eq:application_block_covariance}
\end{equation}
where
\begin{align}
 S_{\rm RT}&=2S_{\rm op}+4(1+\cos^2u)S_{\rm pm},\\
 C_{\rm RT}&=2\cos u\,(S_{\rm op}+4S_{\rm pm}),\\
 S_{\rm NRT}&=2(S_{\rm op}+2S_{\rm pm}).
 \label{eq:application_block_spectra}
\end{align}
The two round trips use the same two one-way streams in different delayed
orders; their cross spectrum must therefore be retained. The exact Relay PSD, $S_U=\boldsymbol w C_z\boldsymbol w^\dagger$, is
\begin{equation}
 S_U=|a|^2\left[2S_{\rm RT}-2\cos(2u)C_{\rm RT}\right]
       +|a|^4S_{\rm NRT}.
 \label{eq:application_SU}
\end{equation}
The equivalent scalar expression is
\begin{align}
 S_U={}&16\sin^2u\,\Big\{
 v\left[\sin^2\frac{3u}{2}+\sin^2\frac{u}{2}+2\sin^2u\right]
 \nonumber\\
 &\hspace{26mm}+4\cos u\sin\frac{3u}{2}\sin\frac{u}{2}S_{\rm pm}\Big\}.
 \label{eq:application_SU_factorized}
\end{align}
For comparison, discarding the off-diagonal block covariance while leaving
the signal response unchanged would give
\begin{equation}
 S_U^{\rm diag}=2|a|^2S_{\rm RT}+|a|^4S_{\rm NRT}.
 \label{eq:application_SU_diagonal}
\end{equation}
The low-frequency expansions of the two spectra are
\begin{align}
 S_U&=u^4(72S_{\rm op}+192S_{\rm pm})
       +O\!\left[u^6(S_{\rm op}+S_{\rm pm})\right],\\
 S_U^{\rm diag}&=16u^2(S_{\rm op}+4S_{\rm pm})
       +O\!\left[u^4(S_{\rm op}+S_{\rm pm})\right].
 \label{eq:application_low_frequency_noise}
\end{align}
Here the frequency-dependent individual noise PSDs are factored out of the
transfer-function expansions. The diagonal approximation misses the leading
secondary-noise cancellation between the two shared round trips.

\subsection{Conditional amplitude and distance information}
\label{sec:binary_information}

The spectra derived above determine the noise weight in a source-amplitude measurement. We consider a local segment in which the detector geometry and binary frequency are effectively constant. The source direction, inclination, polarization, and phase evolution are fixed by conditioning on their values, and the measured signal is written as $h_U(t)=A\,g_U(t)$, where $A>0$ is a dimensionless amplitude scale. Define the transform with respect to ordinary frequency $f$ by
\begin{equation}
 H_U(f)\equiv\int_{-\infty}^{\infty}h_U(t)e^{2\pi ift}\,dt,
 \label{eq:application_Hz_transform}
\end{equation}
with the signal restricted to the analyzed segment. For stationary Gaussian noise with a known, parameter-independent one-sided PSD, the optimal SNR is \citep{Robson2019Sensitivity,LISA-SNR-2021-Babak}
\begin{equation}
 \rho^2=4\int_0^\infty
 \frac{|H_U(f)|^2}{S_U(f)}\,df.
 \label{eq:application_rho}
\end{equation}
The fixed waveform shape implies $\partial h_U/\partial\ln A=h_U$, giving the conditional Fisher identity $\Gamma_{\ln A,\ln A}=\rho^2$. At fixed values of the other waveform parameters, $A\propto d^{-1}$, and the conditional Fisher estimate of the log-distance uncertainty is $\sigma_{\ln d}=\rho^{-1}$. This estimate applies in the high-SNR, local Gaussian regime with the remaining source parameters held fixed \citep{Vallisneri2008Fisher}.

For a narrowband signal with nonzero response, provided that $S_U^{\rm diag}/S_U$ varies negligibly across its bandwidth, the SNR and conditional Fisher-uncertainty ratios obtained with the diagonal noise approximation are
\begin{equation}
 \frac{\rho_{\rm diag}}{\rho}
 =\sqrt{\frac{S_U}{S_U^{\rm diag}}},\qquad
 \frac{\sigma_{\ln d,\rm diag}}{\sigma_{\ln d}}
 =\sqrt{\frac{S_U^{\rm diag}}{S_U}}.
 \label{eq:application_information_ratios}
\end{equation}
Here $\rho_{\rm diag}$ is the SNR computed with $S_U^{\rm diag}$. The same uncertainty ratio holds for $\ln A$. The complex signal response, amplitude, and segment duration cancel in these ratios. The quantity $\sigma_{\ln d,\rm diag}$ is inferred from the curvature of the likelihood constructed with the diagonal noise approximation. A constant PSD rescaling in a single narrowband channel changes this curvature while leaving the maximum-likelihood amplitude estimate and its sampling variance unchanged.

\begin{figure*}[t]
 \centering
 \includegraphics[width=\textwidth]{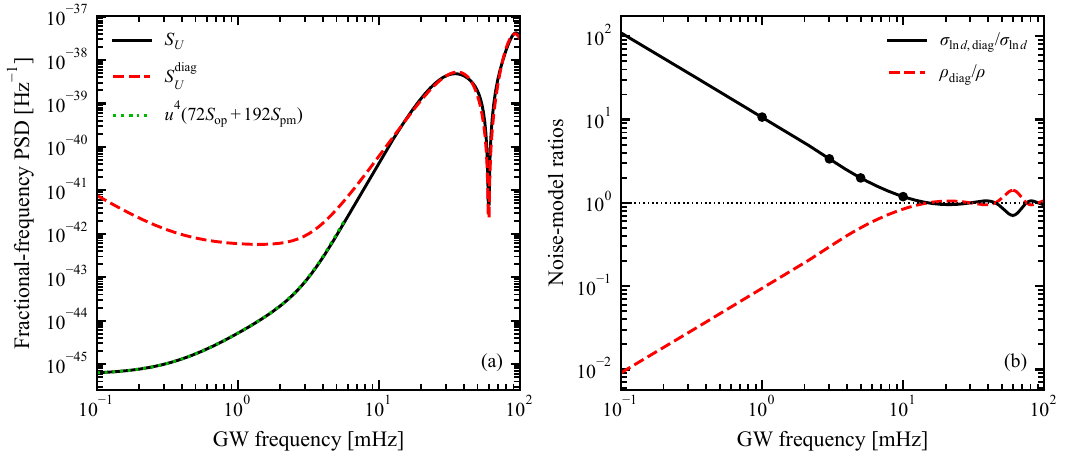}
 \caption{Instrumental noise and conditional information in the equal-arm Relay observable $[U]_2^{16}$ with $L=2.5\times10^9\,{\rm m}$. (a) Comparison of the fractional-frequency PSDs obtained from the full covariance, shown by the black solid curve, the block diagonal approximation, shown by the red dashed curve, and the low-frequency expansion, shown by the green dotted curve. (b) Ratios of $\sigma_{\ln d,\rm diag}/\sigma_{\ln d}$ with the black solid curve and $\rho_{\rm diag}/\rho$ with the red dashed curve for a narrowband signal with the other source parameters fixed. Markers identify the frequencies in Table~\ref{tab:application_binary_ratios}, and the horizontal dotted line denotes unity. The shared transfer zero near $60\,{\rm mHz}$ is masked.}

 \label{fig:application_shared_noise}
\end{figure*}

The PSD ratios are evaluated away from common transfer zeros, and their interpretation as information ratios requires a nonzero measured signal response. The shared zero near $60\,{\rm mHz}$ is masked in Figure~\ref{fig:application_shared_noise}.

We evaluate these expressions with $L=2.5\times10^9\,{\rm m}$ and the
acceleration and optical metrology PSDs specified in Section~\ref{sec:instrument_noise}.
Figure~\ref{fig:application_shared_noise} shows both the spectra and the
information ratios, and Table~\ref{tab:application_binary_ratios} lists four
frequencies relevant to nearly monochromatic compact binaries. At
$1\,{\rm mHz}$ the diagonal approximation overestimates the conditional
Fisher uncertainty by a factor of $10.71$; at $3\,{\rm mHz}$ the factor is $3.37$.
In the acceleration-dominated, long-wavelength limit,
$\rho_{\rm diag}/\rho\simeq\sqrt{3}\,u$.
At higher frequencies the sign of the interference term changes, so
neglecting the covariance can either increase or decrease the nominal error.

\begin{table}[t]
 \centering
 \caption{Conditional Fisher-uncertainty and SNR ratios obtained by omitting the shared-round-trip covariance for a narrowband compact-binary signal in the equal-arm LISA model.}
 \label{tab:application_binary_ratios}
 \begin{tabular}{ccc}
 \hline
 $f\,[{\rm mHz}]$ & $\sigma_{\ln d,\rm diag}/\sigma_{\ln d}$
 & $\rho_{\rm diag}/\rho$\\
 \hline
 1 & 10.709 & 0.09338\\
 3 & 3.373 & 0.29646\\
 5 & 1.991 & 0.50222\\
 10 & 1.187 & 0.84225\\
 \hline
 \end{tabular}
\end{table}

Expanding the geometric blocks in the six one-way streams gives the same Relay spectrum as Eqs.~\eqref{eq:application_SU} and \eqref{eq:application_SU_factorized}. Applying the one-way covariance to the equal-arm Michelson observable recovers the standard spectrum \citep{geometric-tdi-2023-Wang-PRD,LISA-SNR-2021-Babak}
\begin{equation}
 S_{[X]_1^{16}}=64\sin^2u\sin^2(2u)
 \left[S_{\rm op}+2(1+\cos^2u)S_{\rm pm}\right].
 \label{eq:application_X_noise_check}
\end{equation}
The shared-block cross-spectra are needed to recover the noise cancellation present in the one-way TDI combination. Discarding them replaces the physical PSD $S_U$ by $S_U^{\rm diag}$ in the likelihood. Since the Fisher information is weighted by the inverse PSD, this replacement changes the uncertainty predicted by the assumed noise model, while the TDI data and GW response remain the same.

\FloatBarrier
\section{Conclusions}\label{sec5}
We have developed a geometric-link representation of TDI responses using six round-trip and six non-round-trip two-link blocks. The dictionary between these blocks and the one-way measurement streams gives explicit representations of 45 known second-generation combinations while preserving the order of their delay operators. For a fixed, equal-arm constellation, we obtain analytical sky-averaged tensor responses for representative Michelson and Relay observables and long-wavelength limits for the elementary links. The block-noise covariance reproduces the spectrum obtained from direct one-way assembly when the shared-link cross-spectra are retained. For the Relay observable, these cross-spectra contribute to a low-frequency noise cancellation. In the adopted equal-arm LISA instrument model, omitting them overestimates the conditional Fisher uncertainties in compact-binary amplitude or distance by factors of 10.71 and 3.37 at 1 and 3~mHz, respectively, with the other source parameters fixed.

\section*{Acknowledgments}
This work was supported by the National Key R\&D Program of China under Grant 2023YFC2205500, the National Natural Science Foundation of China under Grant 12405060, and the Fundamental Research Funds for the Central Universities under Grant YCJJ20252113.

\paragraph{Data availability.}
This theoretical study uses no observational data. The analytical results are provided in the main text and appendix.

\clearpage
\appendix
\setcounter{equation}{0}
\renewcommand{\theequation}{\Alph{section}\arabic{equation}}
\section{Geometric-link representations of the TDI catalogue}\label{appendixA}
This appendix collects the representations of the 45 combinations considered in \citet{geometric-tdi-2022-Wang-PRD}. The superscripts retain the link counts and the subscripts retain the original enumeration. The $RT$ and $NRT$ observables are defined in Eqs.~\eqref{rt} and \eqref{nrt}. Residual one-way streams are shown explicitly whenever a complete pairing is not used. Barred indices follow the inverse-delay convention in Section~\ref{sec2}. Each expression can be expanded using the mapping rules without changing the order of the delay operators.
\subsection{12-link combinations}\label{12-LINKS}
\begin{equation}
\begin{aligned}
 {}[\alpha]_{1}^{12} &= \left(1 - {D}_{2'1'3'} \right) \eta_1 + (-1 + {D}_{312}) \eta_{1'} + \left( -{D}_{2'1'3'3} + {D}_3 \right) \eta_2\\
 &\quad + \left( - {D}_{2'1'} + {D}_{3122'1'} \right) \eta_{2'} + \left( -{D}_{2'1'3'31} + {D}_{31} \right) \eta_3\\
 &\quad + \left( -{D}_{2'} + {D}_{3122'} \right) \eta_{3'}
\end{aligned}
\label{app:catalogue-1}
\end{equation}

\begin{equation}
\begin{aligned}
 {}[\alpha]_2^{12} &= -D_{311'3'\bar{2}}D_{\bar{1}}NRT_{1\rightarrow3\rightarrow2} + D_{2'1'3'\bar{2}}D_{\bar{2}'}RT_{1\rightarrow3\rightarrow1}\\
 &\quad + (D_{31}-D_{2'})NRT_{1\rightarrow2\rightarrow3} + NRT_{3\rightarrow2\rightarrow1} -D_{2'1'3'\bar{2}\bar{2'}}\eta_{1} - \eta_{1'}
\end{aligned}
\label{app:catalogue-2}
\end{equation}

\begin{equation}
\begin{aligned}
 {}[\alpha]_3^{12} &= -D_{2'\bar{1}3'\bar{2}}RT_{3\rightarrow2\rightarrow3} + (1-D_{33'\bar{2}1'\bar{3}})(\eta_{1} - \eta_{1'}) + D_{2'\bar{1}}\eta_{2}\\
 &\quad + (D_{3}-D_{2'\bar{1}})\eta_{2'} + (D_{2'\bar{1}3'\bar{2}}-D_{33'\bar{2}})\eta_{3} + D_{33'\bar{2}}\eta_{3'}
\end{aligned}
\label{app:catalogue-3}
\end{equation}

\subsection{14-link combinations}\label{14-LINKS}
\begin{equation}
\begin{aligned}
 {}[U]_1^{14} &= -D_{2'1'3'\bar{2}\bar{1}}RT_{2\rightarrow1\rightarrow2} + D_{3}NRT_{3\rightarrow1\rightarrow2}\\
 &\quad + (D_{2'1'3'\bar{2}}-D_{33'2'1'3'\bar{2}})D_{\bar{1}}NRT_{1\rightarrow3\rightarrow2}\\
 &\quad +(D_{33'2'}-D_{2'})NRT_{1\rightarrow2\rightarrow3} + \eta_{1} - \eta_{1'}
\end{aligned}
\label{app:catalogue-4}
\end{equation}

\begin{equation}
\begin{aligned}
 {}[U]_2^{14} &= (1- D_{3122'\bar{1}}D_{\bar{3}})NRT_{3\rightarrow2\rightarrow1} + D_{31}RT_{3\rightarrow1\rightarrow3}\\
 &\quad - D_{2'1'3'}RT_{1\rightarrow3\rightarrow1} + (D_{2'1'3'2'2\bar{3}'}D_{\bar{1}'}-D_{2'})NRT_{1\rightarrow2\rightarrow3}\\
 &\quad + (D_{3122'\bar{1}\bar{3}}-1)\eta_{1'}
\end{aligned}
\label{app:catalogue-5}
\end{equation}

\begin{equation}
\begin{aligned}
 {}[EP]_1^{14} &= D_{3}RT_{2\rightarrow3\rightarrow2} -D_{2'1'3'\bar{2}1'}NRT_{1\rightarrow3\rightarrow2}\\
 &\quad + (D_{311'\bar{3}2'}-D_{2'})NRT_{1\rightarrow2\rightarrow3} + (1-D_{311'\bar{3}})(\eta_{1}-\eta_{1'})\\
 &\quad + D_{2'1'3'\bar{2}}(\eta_{3}-\eta_{3'})
\end{aligned}
\label{app:catalogue-6}
\end{equation}

\begin{equation}
\begin{aligned}
 {}[EP]_2^{14} &= -D_{2'\bar{1}3'\bar{2}1'}RT_{2\rightarrow1\rightarrow2} +(\eta_{1}-\eta_{1'}) + (D_{2'\bar{1}}-D_{33'2'\bar{1}})(\eta_{2}-\eta_{2'})\\
 &\quad + (D_{2'\bar{1}3'\bar{2}}-D_{33'2'\bar{1}3'\bar{2}})(\eta_{3}-\eta_{3'})
\end{aligned}
\label{app:catalogue-7}
\end{equation}

\subsection{16-link Relay, Michelson and Sagnac combinations}\label{16-LINKS}
\begin{equation}
\begin{aligned}
 {}[U]_1^{16} &= (1-D_{2'1'3'})RT_{1\rightarrow2\rightarrow1}\\
 &\quad + (-1-D_{2'1'3'33'}+D_{33'}+D_{33'2'1'\bar{3}})NRT_{2\rightarrow3\rightarrow1}\\
 &\quad + (D_{2'1'3'33'2'1'\bar{3}}-D_{33'2'1'\bar{3}})\eta_{1}\\
 &\quad + (D_{33'2'1'\bar{3}2'1'}-D_{2'1'})\eta_{2'}
\end{aligned}
\label{app:catalogue-8}
\end{equation}

\begin{equation}
\begin{aligned}
 {}[U]_2^{16} &= (1-D_{2'1'3'3\bar{1}'\bar{2}'})RT_{1\rightarrow2\rightarrow1}\\
 &\quad + (D_{33'2'1'}-D_{2'1'})RT_{2\rightarrow1\rightarrow2} + (D_{33'}-1)NRT_{2\rightarrow3\rightarrow1}\\
 &\quad + (D_{2'1'3'3\bar{1}'}-D_{33'2'1'3'3\bar{1}'})D_{\bar{2}'}NRT_{2\rightarrow3\rightarrow1}
\end{aligned}
\label{app:catalogue-9}
\end{equation}

\begin{equation}
\begin{aligned}
 {}[U]_3^{16} &= (1-D_{33'2'1'\bar{3}\bar{3}'}D_{\bar{3}})RT_{1\rightarrow2\rightarrow1} - D_{33'2'1'\bar{3}\bar{3}'}RT_{2\rightarrow1\rightarrow2}\\
 &\quad + (D_{33'}-D_{2'1'3'}-1)NRT_{2\rightarrow3\rightarrow1} + D_{33'2'1'\bar{3}\bar{3}'\bar{3}}\eta_{1'}\\
 &\quad - D_{33'2'1'\bar{3}}\eta_{1} - D_{2'1'}\eta_{2'} + D_{2'1'3'2'1'\bar{3}\bar{3}'\bar{1}'}\eta_{3'}
\end{aligned}
\label{app:catalogue-10}
\end{equation}

\begin{equation}
\begin{aligned}
 {}[U]_4^{16} &= (1+D_{33'2'1'12})RT_{1\rightarrow2\rightarrow1} + (-D_{2'1'}-D_{2'1'3'})RT_{2\rightarrow1\rightarrow2}\\
 &\quad + (D_{33'}-1)NRT_{2\rightarrow3\rightarrow1}\\
 &\quad + (D_{33'2'1'}-D_{2'1'3'33'3})NRT_{2\rightarrow3\rightarrow1}
\end{aligned}
\label{app:catalogue-11}
\end{equation}

\begin{equation}
\begin{aligned}
 {}[U]_5^{16} &= (1+D_{2'1'3'312\bar{3}'}D_{\bar{3}})RT_{1\rightarrow2\rightarrow1} + (D_{33'}-1)NRT_{2\rightarrow3\rightarrow1}\\
 &\quad - (D_{2'1'}+D_{33'2'1'\bar{3}}D_{\bar{3}'})RT_{2\rightarrow1\rightarrow2}\\
 &\quad + (D_{33'2'1'\bar{3}\bar{3}'}-D_{2'1'3'3})NRT_{1\rightarrow3\rightarrow2}
\end{aligned}
\label{app:catalogue-12}
\end{equation}

\begin{equation}
\begin{aligned}
 {}[U]_6^{16} &= (1+D_{33'2'2})RT_{1\rightarrow2\rightarrow1} - (D_{2'1'}+D_{2'1'3'3\bar{1}'\bar{1}})RT_{2\rightarrow1\rightarrow2}\\
 &\quad + D_{33'}RT_{1\rightarrow3\rightarrow1} - D_{33'2'233'\bar{2}}NRT_{1\rightarrow3\rightarrow2} - NRT_{2\rightarrow3\rightarrow1}\\
 &\quad + D_{2'1'3'3\bar{1}'}RT_{2\rightarrow3\rightarrow2}
\end{aligned}
\label{app:catalogue-13}
\end{equation}

\begin{equation}
\begin{aligned}
 {}[X]_{1}^{16} &= (D_{33'2'22'2}-D_{2'233'})RT_{1\rightarrow2\rightarrow1} +(1-D_{2'2}) RT_{1\rightarrow2\rightarrow1}\\
 &\quad +(D_{33'2'2}-D_{2'233'33'})RT_{1\rightarrow3\rightarrow1} +(D_{33'}-1)RT_{1\rightarrow3\rightarrow1}
\end{aligned}
\label{app:catalogue-14}
\end{equation}

\begin{equation}
\begin{aligned}
 {}[X]_{2}^{16} &= (1-D_{2'2})RT_{1\rightarrow2\rightarrow1} +(D_{33'2'2}-D_{2'233'\bar{2}\bar{2}'}) RT_{1\rightarrow2\rightarrow1}\\
 &\quad +(D_{33'}-1)RT_{1\rightarrow3\rightarrow1}\\
 &\quad +(D_{2'233'\bar{2}}D_{\bar{2}'}-D_{33'2'233'\bar{2}}D_{\bar{2}'})RT_{1\rightarrow3\rightarrow1}
\end{aligned}
\label{app:catalogue-15}
\end{equation}

\begin{equation}
\begin{aligned}
 {}[T]_{1}^{16} &= -D_{33'2'\bar{1}}NRT_{3\rightarrow2\rightarrow1} + (D_{3}-D_{2'1'3'\bar{2}\bar{1}})NRT_{3\rightarrow1\rightarrow2}\\
 &\quad + D_{2'1'3'\bar{2}}NRT_{1\rightarrow3\rightarrow2} + (D_{33'2'\bar{1}\bar{3}2'}-D_{2'})NRT_{1\rightarrow2\rightarrow3} + \eta_{1}\\
 &\quad + (D_{33'2'\bar{1}\bar{3}}-1)\eta_{1'} - D_{2'1'3'\bar{2}\bar{1}3'2'}\eta_{3}
\end{aligned}
\label{app:catalogue-16}
\end{equation}

\begin{equation}
\begin{aligned}
 {}[T]_{2}^{16} &= (D_{3}-D_{2'1'3'\bar{2}\bar{1}})RT_{2\rightarrow1\rightarrow2}\\
 &\quad + (D_{2'1'3'\bar{2}}-D_{33'33'\bar{2}})D_{\bar{1}}NRT_{1\rightarrow3\rightarrow2} - D_{2'}NRT_{1\rightarrow2\rightarrow3}\\
 &\quad + (D_{33'33'\bar{2}\bar{1}\bar{3}}-1)(\eta_{1}-\eta_{1'}) + D_{33'3}\eta_{2'} + D_{2'1'3'\bar{2}\bar{1}3'3\bar{1}'}\eta_{3'}
\end{aligned}
\label{app:catalogue-17}
\end{equation}

\begin{equation}
\begin{aligned}
 {}[T]_{3}^{16} &= (1-D_{2'1'3'\bar{2}1'3'})NRT_{3\rightarrow2\rightarrow1}\\
 &\quad + (-D_{2'}-D_{2'1'3'\bar{2}}+D_{31}+D_{311'3'2'})NRT_{1\rightarrow2\rightarrow3}\\
 &\quad + (D_{311'3'}-1)\eta_{1'} + (D_{2'1'3'\bar{2}}-D_{311'3'2'1'3'\bar{2}})\eta_{3}
\end{aligned}
\label{app:catalogue-18}
\end{equation}

\begin{equation}
\begin{aligned}
 {}[T]_{4}^{16} &= -D_{311'3'\bar{2}1'\bar{3}}D_{\bar{3}'}RT_{2\rightarrow1\rightarrow2} - NRT_{2\rightarrow3\rightarrow1}\\
 &\quad + D_{2'1'11'\bar{3}\bar{2}}D_{\bar{2}'}RT_{1\rightarrow3\rightarrow1} + D_{311'}\eta_{2'}\\
 &\quad + (D_{2'1'}+D_{3})RT_{2\rightarrow3\rightarrow2} + (1+D_{2'1'11'\bar{3}}-D_{2'1'11'\bar{3}\bar{2}\bar{2}'})\eta_{1}\\
 &\quad + D_{311'3'\bar{2}}(\eta_{3'}-\eta_{3})
\end{aligned}
\label{app:catalogue-19}
\end{equation}

\begin{equation}
\begin{aligned}
 {}[T]_{5}^{16} &= -D_{2'1'3'\bar{2}}RT_{3\rightarrow2\rightarrow3} + (D_{3}+D_{311'})RT_{2\rightarrow3\rightarrow2}\\
 &\quad - D_{311'11'3'\bar{2}}D_{\bar{1}}NRT_{1\rightarrow3\rightarrow2} - D_{2'}NRT_{1\rightarrow2\rightarrow3}\\
 &\quad + (1-D_{2'1'3'\bar{2}1'1\bar{2}'})(\eta_{1}-\eta_{1'}) + D_{311'11'}\eta_{2'} + D_{2'1'3'\bar{2}}\eta_{3}
\end{aligned}
\label{app:catalogue-20}
\end{equation}

\begin{equation}
\begin{aligned}
 {}[T]_{6}^{16} &= (-1-D_{2'1'3'}+D_{3122'\bar{3}}+D_{312})NRT_{2\rightarrow3\rightarrow1}\\
 &\quad + (D_{3}-D_{2'1'3'2'1'})NRT_{1\rightarrow3\rightarrow2} + (1-D_{3122'1'\bar{3}})\eta_{1}\\
 &\quad + (D_{2'1'3'2'1'1'2\bar{3}'})\eta_{2'}
\end{aligned}
\label{app:catalogue-21}
\end{equation}

\begin{equation}
\begin{aligned}
 {}[T]_{7}^{16} &= NRT_{3\rightarrow2\rightarrow1} - D_{2'1'3'2'}NRT_{2\rightarrow1\rightarrow3} - D_{2'1'}NRT_{3\rightarrow1\rightarrow2}\\
 &\quad + (D_{31}+D_{3122'})RT_{3\rightarrow1\rightarrow3} - D_{2'1'3'2'23}NRT_{1\rightarrow3\rightarrow2}\\
 &\quad + D_{3122'22'}NRT_{1\rightarrow2\rightarrow3}
\end{aligned}
\label{app:catalogue-22}
\end{equation}

\begin{equation}
\begin{aligned}
 {}[T]_{8}^{16} &= NRT_{3\rightarrow2\rightarrow1} - NRT_{2\rightarrow3\rightarrow1} - D_{2'1'}NRT_{3\rightarrow1\rightarrow2}\\
 &\quad + D_{2'1'3'2'\bar{1}}D_{\bar{1}'}RT_{3\rightarrow2\rightarrow3} + D_{31}RT_{3\rightarrow1\rightarrow3}\\
 &\quad - D_{3122'\bar{1}3'\bar{2}}D_{\bar{1}}NRT_{1\rightarrow3\rightarrow2} + D_{2'1'3'2'\bar{1}\bar{1}'\bar{2}'}(\eta_{1'}-\eta_{1})\\
 &\quad + D_{3122'\bar{1}}(\eta_{2'}-\eta_{2})
\end{aligned}
\label{app:catalogue-23}
\end{equation}

\begin{equation}
\begin{aligned}
 {}[T]_{9}^{16} &= -D_{2'}RT_{3\rightarrow1\rightarrow3} + D_{2'22'\bar{1}3'\bar{2}}D_{\bar{2}'}RT_{1\rightarrow3\rightarrow1}\\
 &\quad + D_{33'3\bar{1}'\bar{1}\bar{3}}NRT_{2\rightarrow3\rightarrow1} - D_{33'3\bar{1}'}D_{\bar{1}}RT_{2\rightarrow3\rightarrow2}\\
 &\quad + (1-D_{2'22'\bar{1}3'\bar{2}\bar{2}'}+D_{33'}-D_{33'3\bar{1}'\bar{1}\bar{3}})\eta_{1} - \eta_{1'} + D_{2'22'\bar{1}}\eta_{2}\\
 &\quad + (D_{3}-D_{2'22'\bar{1}})\eta_{2'}
\end{aligned}
\label{app:catalogue-24}
\end{equation}

\begin{equation}
\begin{aligned}
 {}[T]_{11}^{16} &= RT_{1\rightarrow2\rightarrow1} + (D_{2'\bar{1}3'2'1'\bar{3}}-D_{33'2'1'\bar{3}})D_{\bar{2}}NRT_{2\rightarrow1\rightarrow3}\\
 &\quad - D_{33'2'1'\bar{3}\bar{2}\bar{1}}D_{\bar{3}}NRT_{3\rightarrow2\rightarrow1} + (D_{33'}-D_{2'\bar{1}3'})NRT_{2\rightarrow3\rightarrow1}\\
 &\quad + (D_{2'\bar{1}3'2'1'\bar{3}\bar{2}\bar{2'}}-1)\eta_{1} + D_{2'\bar{1}}\eta_{2} + (D_{3}-D_{2'\bar{1}})\eta_{2'}
\end{aligned}
\label{app:catalogue-25}
\end{equation}

\begin{equation}
\begin{aligned}
 {}[T]_{12}^{16} &= -D_{2'\bar{1}3'\bar{2}1'3'}RT_{1\rightarrow3\rightarrow1} + D_{33'}NRT_{2\rightarrow3\rightarrow1}\\
 &\quad - D_{2'\bar{1}3'\bar{2}}NRT_{1\rightarrow2\rightarrow3} + (1-D_{33'2'\bar{1}\bar{3}})(\eta_{1}-\eta_{1'})\\
 &\quad + (D_{2'\bar{1}}-D_{33'2'1'\bar{3}2'\bar{1}})\eta_{2} - (D_{3}+D_{2'\bar{1}}-D_{33'2'1'\bar{3}2'\bar{1}})\eta_{2'}\\
 &\quad + D_{2'\bar{1}3'\bar{2}}\eta_{3}
\end{aligned}
\label{app:catalogue-26}
\end{equation}

\begin{equation}
\begin{aligned}
 {}[T]_{14}^{16} &= -D_{2'}RT_{3\rightarrow1\rightarrow3} + D_{3}NRT_{3\rightarrow1\rightarrow2} + D_{2'22'\bar{1}\bar{1}'}D_{\bar{1}}RT_{2\rightarrow3\rightarrow2}\\
 &\quad - D_{33'2'\bar{1}\bar{3}2'\bar{1}}D_{\bar{1}'}RT_{3\rightarrow2\rightarrow3} + (1-D_{33'2'\bar{1}})(\eta_{1}-\eta_{1'})\\
 &\quad + (D_{2'22'\bar{1}}-D_{33'2'\bar{1}})\eta_{2} - D_{2'22'\bar{1}\bar{1}'\bar{1}}\eta_{2'}\\
 &\quad + D_{2'22'\bar{1}\bar{1}'\bar{1}3'\bar{2}}\eta_{3}
\end{aligned}
\label{app:catalogue-27}
\end{equation}

\begin{equation}
\begin{aligned}
 {}[T]_{15}^{16} &= (D_{3}-D_{2'\bar{1}3'\bar{2}1'})RT_{2\rightarrow1\rightarrow2} + (1-D_{33'33'\bar{2}1'\bar{3}})(\eta_{1}-\eta_{1'})\\
 &\quad + (D_{2'\bar{1}}-D_{2'\bar{1}3'\bar{2}1'33'})\eta_{2} + (D_{33'3}-D_{2'\bar{1}})\eta_{2'}\\
 &\quad + (D_{2'\bar{1}3'\bar{2}}-D_{33'33'\bar{2}})(\eta_{3}-\eta_{3'})
\end{aligned}
\label{app:catalogue-28}
\end{equation}

\begin{equation}
\begin{aligned}
 {}[T]_{16}^{16} &= RT_{1\rightarrow2\rightarrow1} + D_{33'}NRT_{3\rightarrow2\rightarrow1} - D_{33'31\bar{2}'\bar{3}'\bar{3}}D_{\bar{2}}NRT_{2\rightarrow1\rightarrow3}\\
 &\quad + D_{2'1'1\bar{2}'\bar{2}}D_{\bar{2}'}RT_{1\rightarrow3\rightarrow1} - D_{33'31\bar{2}'}D_{\bar{3}'}NRT_{3\rightarrow1\rightarrow2}\\
 &\quad - D_{2'}RT_{3\rightarrow2\rightarrow3} - D_{2'1'1\bar{2}'\bar{2}\bar{2'}}\eta_{1} + (D_{2'1'1\bar{2}'}-1)\eta_{1'}\\
 &\quad + D_{2'1'1\bar{2}'\bar{2}\bar{2}'3\bar{1}'}\eta_{3'}
\end{aligned}
\label{app:catalogue-29}
\end{equation}

\begin{equation}
\begin{aligned}
 {}[T]_{18}^{16} &= RT_{1\rightarrow2\rightarrow1} + D_{2'1'\bar{3}}D_{\bar{2}}NRT_{2\rightarrow1\rightarrow3} - D_{33'\bar{2}1'\bar{3}}D_{\bar{3}'}RT_{2\rightarrow1\rightarrow2}\\
 &\quad - NRT_{2\rightarrow3\rightarrow1} - D_{2'1'\bar{3}\bar{2}}RT_{3\rightarrow2\rightarrow3} + D_{33'\bar{2}1'\bar{3}\bar{3}'}RT_{2\rightarrow3\rightarrow2}\\
 &\quad + D_{2'1'\bar{3}\bar{2}1'1\bar{2}'}(\eta_{1'}-\eta_{1}) + D_{33'\bar{2}}(\eta_{3'}-\eta_{3})
\end{aligned}
\label{app:catalogue-30}
\end{equation}

\begin{equation}
\begin{aligned}
 {}[T]_{19}^{16} &= -D_{2'\bar{1}}RT_{1\rightarrow2\rightarrow1} + D_{33'\bar{2}\bar{1}}NRT_{3\rightarrow1\rightarrow2} + D_{2'\bar{1}3'3\bar{1}'}D_{\bar{1}}RT_{2\rightarrow3\rightarrow2}\\
 &\quad - D_{33'\bar{2}}D_{\bar{1}}NRT_{1\rightarrow3\rightarrow2} - D_{33'\bar{2}\bar{1}3'2'\bar{1}}D_{\bar{1}'}RT_{3\rightarrow2\rightarrow3} + (\eta_{1}-\eta_{1'})\\
 &\quad + D_{2'\bar{1}}\eta_{2} + (D_{3}- D_{2'\bar{1}3'3\bar{1}'\bar{1}})\eta_{2'} + D_{2'\bar{1}3'3\bar{1}'\bar{1}3'\bar{2}}\eta_{3}
\end{aligned}
\label{app:catalogue-31}
\end{equation}

\subsection{16-link Beacon, Monitor and mixed combinations}\label{16-LINKS-PET}
\begin{equation}
\begin{aligned}
 {}[P]_{1}^{16} &= (1-D_{2'\bar{1}3'31\bar{2}'})RT_{1\rightarrow2\rightarrow1} + (D_{33'2'}-D_{2'\bar{1}})RT_{2\rightarrow1\rightarrow2}\\
 &\quad + (D_{2'\bar{1}3'31\bar{2}'}+D_{33'}-D_{33'2'\bar{1}3'31\bar{2}'}-1)\eta_{1'}\\
 &\quad + (D_{2'\bar{1}}-D_{2'\bar{1}3'3}-D_{33'2'\bar{1}}+D_{33'2'\bar{1}3'3})\eta_{2}
\end{aligned}
\label{app:catalogue-32}
\end{equation}

\begin{equation}
\begin{aligned}
 {}[P]_{2}^{16} &= -D_{2'\bar{1}}RT_{2\rightarrow1\rightarrow2} + (D_{2'\bar{1}3'33'2'\bar{1}}-D_{33'2'\bar{1}})D_{\bar{3}}NRT_{3\rightarrow2\rightarrow1}\\
 &\quad + (D_{3}-D_{2'\bar{1}3'3})NRT_{3\rightarrow1\rightarrow2} + \eta_{1} + (D_{33'2'\bar{1}\bar{3}}-1)\eta_{1'}\\
 &\quad + (D_{2'\bar{1}}-D_{33'2'\bar{1}\bar{3}2'\bar{1}})\eta_{2} + D_{33'2'\bar{1}\bar{3}2'\bar{1}}\eta_{2'}
\end{aligned}
\label{app:catalogue-33}
\end{equation}

\begin{equation}
\begin{aligned}
 {}[E]_1^{16} &= ({D}_{311'\bar{3}} - 1) \left[D_{2'}RT_{3\rightarrow2\rightarrow3} - ({D}_{2'1'1\bar{2}'} - 1) \eta_{1'}\right]\\
 &\quad -({D}_{2'1'1\bar{2}'} - 1)\left[D_{3}RT_{2\rightarrow3\rightarrow2} - ({D}_{311'\bar{3}} - 1) \eta_{1}\right]
\end{aligned}
\label{app:catalogue-34}
\end{equation}

\begin{equation}
\begin{aligned}
 {}[E]_{2}^{16} &= (-D_{2'}+D_{31})RT_{3\rightarrow2\rightarrow3} - D_{311'1\bar{2}'3\bar{1}'}D_{\bar{1}}RT_{2\rightarrow3\rightarrow2}\\
 &\quad + (1-D_{2'1'1\bar{2}'}-D_{2'1'1\bar{2}'3\bar{1}'\bar{2}'}+D_{311'1\bar{2}'})(\eta_{1}-\eta_{1'}) + D_{3}\eta_{2}\\
 &\quad + D_{2'1'1\bar{2}'3\bar{1}'}\eta_{3'}
\end{aligned}
\label{app:catalogue-35}
\end{equation}

\begin{equation}
\begin{aligned}
 {}[PE]_{1}^{16} &= (-D_{2'}-D_{311'3'\bar{2}\bar{1}}D_{\bar{1}'})RT_{3\rightarrow2\rightarrow3}\\
 &\quad + (D_{2'1'1\bar{2}'3\bar{1}'}D_{\bar{1}}+D_{3})RT_{2\rightarrow3\rightarrow2} + (1-D_{2'1'1\bar{2}'})(\eta_{1}-\eta_{1'})\\
 &\quad + (D_{311'}-D_{2'1'1\bar{2}'3\bar{1}'\bar{1}})\eta_{2} + (D_{311'3'\bar{2}\bar{1}\bar{1}'}-D_{311'3'\bar{2}})\eta_{3}
\end{aligned}
\label{app:catalogue-36}
\end{equation}

\begin{equation}
\begin{aligned}
 {}[PE]_{2}^{16} &= -D_{311'\bar{3}}D_{\bar{3}'}RT_{2\rightarrow1\rightarrow2} + D_{2'1'1\bar{2}'}D_{\bar{2}}RT_{3\rightarrow1\rightarrow3}\\
 &\quad - (D_{2'}+D_{2'1'1\bar{2}'\bar{2}})RT_{3\rightarrow2\rightarrow3} + (D_{3}+D_{311'\bar{3}\bar{3}'})RT_{2\rightarrow3\rightarrow2}\\
 &\quad + (\eta_{1}-\eta_{1'}) + D_{311'\bar{3}\bar{3}'11'}\eta_{2'} - D_{2'1'1\bar{2}'\bar{2}1'1}\eta_{3}
\end{aligned}
\label{app:catalogue-37}
\end{equation}

\begin{equation}
\begin{aligned}
 {}[PE]_{3}^{16} &= - D_{2'}RT_{3\rightarrow2\rightarrow3} + (D_{3}+D_{311'})RT_{2\rightarrow3\rightarrow2}\\
 &\quad - D_{311'11'3'\bar{2}}D_{\bar{1}}NRT_{1\rightarrow3\rightarrow2} -D_{2'1'1\bar{2}}RT_{1\rightarrow2\rightarrow1} + \eta_{1}\\
 &\quad + (D_{2'1'1\bar{2}'}-1)\eta_{1'} + D_{311'11'}\eta_{2'} + D_{2'1'1\bar{2}'33'\bar{2}}(\eta_{3}-\eta_{3'})
\end{aligned}
\label{app:catalogue-38}
\end{equation}

\begin{equation}
\begin{aligned}
 {}[PE]_{4}^{16} &= (1+D_{33'2'2})RT_{1\rightarrow2\rightarrow1} - (D_{2'\bar{1}}+D_{2'\bar{1}3'311'})RT_{2\rightarrow1\rightarrow2}\\
 &\quad + D_{33'}RT_{1\rightarrow3\rightarrow1} - D_{2'\bar{1}3'3}RT_{2\rightarrow3\rightarrow2} - \eta_{1'} + D_{2'\bar{1}}\eta_{2}\\
 &\quad - D_{33'2'233'\bar{2}}(\eta_{3}-\eta_{3'})
\end{aligned}
\label{app:catalogue-39}
\end{equation}

\begin{equation}
\begin{aligned}
 {}[PE]_{5}^{16} &= (1+D_{2'1'123\bar{3}'}D_{\bar{3}})RT_{1\rightarrow2\rightarrow1} + (D_{33'}-1)NRT_{2\rightarrow3\rightarrow1}\\
 &\quad - (D_{33'2'1'\bar{3}}+D_{33'2'1'\bar{3}\bar{3}'\bar{3}})D_{\bar{3}'}RT_{2\rightarrow1\rightarrow2}\\
 &\quad + (D_{2'1'12\bar{3}'\bar{3}\bar{2}}D_{\bar{1}}-D_{2'1'})NRT_{1\rightarrow3\rightarrow2}
\end{aligned}
\label{app:catalogue-40}
\end{equation}

\begin{equation}
\begin{aligned}
 {}[PE]_{6}^{16} &= (1+D_{2'\bar{1}3'312\bar{2}'}D_{\bar{3}})RT_{1\rightarrow2\rightarrow1}\\
 &\quad - (D_{2'\bar{1}}+D_{33'2'1'\bar{3}}D_{\bar{3}'})RT_{2\rightarrow1\rightarrow2} + D_{33'}NRT_{2\rightarrow3\rightarrow1}\\
 &\quad - D_{2'\bar{1}3'3}NRT_{1\rightarrow3\rightarrow2} -\eta_{1'} + D_{2'\bar{1}}\eta_{2} + D_{33'2'1'\bar{3}\bar{3}'\bar{1}'}\eta_{3}\\
 &\quad - D_{33'2'1'\bar{3}\bar{3}'\bar{1}'}\eta_{3'}
\end{aligned}
\label{app:catalogue-41}
\end{equation}

\begin{equation}
\begin{aligned}
 {}[PE]_{7}^{16} &= (1+D_{2'1'3'3\bar{1}'2\bar{3}'}D_{\bar{3}})RT_{1\rightarrow2\rightarrow1} - D_{2'1'}RT_{2\rightarrow1\rightarrow2}\\
 &\quad - NRT_{2\rightarrow3\rightarrow1} + D_{33'2'\bar{1}\bar{3}\bar{3}'}NRT_{1\rightarrow3\rightarrow2} + D_{33'}\eta_{1'} - D_{33'2'\bar{1}}\eta_{2}\\
 &\quad - D_{2'1'3'3\bar{1}'}\eta_{3} + D_{2'1'3'3\bar{1}'}\eta_{3'}
\end{aligned}
\label{app:catalogue-42}
\end{equation}

\begin{equation}
\begin{aligned}
 {}[PE]_{8}^{16} &= (1+D_{33'2'\bar{1}\bar{1}'2})RT_{1\rightarrow2\rightarrow1} - (D_{2'\bar{1}}+D_{2'\bar{1}3'3})RT_{2\rightarrow1\rightarrow2}\\
 &\quad + (D_{33'}-1)\eta_{1'} + D_{2'\bar{1}}\eta_{2} + D_{2'\bar{1}3'33'3\bar{1}'}\eta_{3'}\\
 &\quad + (D_{33'2'\bar{1}\bar{1}'}-D_{2'\bar{1}3'33'3\bar{1}'})\eta_{3}
\end{aligned}
\label{app:catalogue-43}
\end{equation}

\begin{equation}
\begin{aligned}
 {}[PE]_{9}^{16} &= (1+D_{33'})RT_{1\rightarrow2\rightarrow1} - (D_{2'\bar{1}}+D_{2'\bar{1}3'3})RT_{2\rightarrow1\rightarrow2}\\
 &\quad + D_{2'\bar{1}3'33'3\bar{1}'}D_{\bar{2}'}NRT_{2\rightarrow3\rightarrow1} - \eta_{1'} + D_{2'\bar{1}}\eta_{2}\\
 &\quad + (D_{33'33'\bar{2}\bar{1}\bar{1}'}-D_{33'33'\bar{2}})\eta_{3}
\end{aligned}
\label{app:catalogue-44}
\end{equation}

\begin{equation}
\begin{aligned}
 {}[PE]_{10}^{16} &= (1+D_{33'})RT_{1\rightarrow2\rightarrow1}\\
 &\quad - (D_{2'1'1\bar{2}'\bar{2}\bar{1}}+D_{33'33'\bar{2}1'\bar{3}}D_{\bar{3}'})RT_{2\rightarrow1\rightarrow2}\\
 &\quad - D_{2'1'1\bar{2}'}D_{\bar{2}}RT_{3\rightarrow1\rightarrow3} - NRT_{2\rightarrow3\rightarrow1} + (-D_{2'}\eta_{3'}+\eta_{1'})\\
 &\quad + (-D_{2'1'}+D_{2'1'1\bar{2}'\bar{2}\bar{1}})\eta_{2} - D_{33'33'\bar{2}}\eta_{3} + D_{33'33'\bar{2}}\eta_{3'}
\end{aligned}
\label{app:catalogue-45}
\end{equation}

\clearpage
\bibliographystyle{numericrefs}
\bibliography{references}
\end{document}